\documentclass[aps, preprint, groupedaddress, floatfix]{revtex4}

\usepackage{epsfig}
\usepackage{graphicx}
\usepackage{amsmath}

\newcommand{\lsm}{La$_{1-x}$Sr$_{x}$MnO$_3$}

\begin{document}

\title{Ferroelectric control of magnetization in La$_{1-x}$Sr$_x$MnO$_3$ manganites: A first-principles study}

\author{Hanghui~Chen$^{1,3}$ and Sohrab Ismail-Beigi$^{1,2,3}$}

\affiliation{
 $^1$Department of Physics, Yale University, New Haven,
Connecticut, 06511, USA\\
 $^2$Department of Applied Physics, Yale
University, New Haven, Connecticut, 06511, USA\\
 $^3$Center for Research on Interface Structures
and Phenomena (CRISP), Yale University, New Haven, CT 06511, USA}

\date{\today}

\begin{abstract}

We present a first-principles study on the interface between 
perovskite ferroelectrics (PbTiO$_3$)
and conducting magnetic manganites (\lsm). 
We show that by switching the ferroelectric 
polarization, additional carriers are accumulated or depleted 
at the interfacial region of the manganite and that this change
in carrier density can modify the magnetic spin configuration of the 
interfacial Mn, which is  consistent with the 
experimentally observed anomalously large change in the magnetization. 
We also describe an unexpected purely interfacial phenomenon whereby the 
ferroelectric polarization of the interfacial region changes the magnetic
energetics --- a degree of freedom not present in bulk manganites.
Theoretically, we show the the ground-state magnetic structure depends
sensitively on the precise choice of Hubbard $U$ parameter within the
widely-used DFT+$U$ class of exchange correlation functionals. 
We provide a simple Ising-like model that explains the evolution of
the magnetic structure with $U$ in tandem with a discussion of various
different ways in which one might try to choose an appropriate $U$ parameter.

\end{abstract}

\maketitle

\section{Introduction}

Multiferroics have been one of the most intensively studied materials
during the past decade \cite{Eerenstein-Nature-2006,
  Ramesh-NatMat-2007}.  The coexistence of more than one order
parameter in a single phase and their coupling may open new routes to
the next generation of electronic devices. For instance, the
possibility of controlling magnetization via external electric fields
may find promising applications in spintronics. The origin of
magnetoelectric multiferroicity lies in a nonzero magnetoelectric
coupling which may occur due to many different mechanisms (for recent
reviews, see \cite{Wang-AdvPhys-2009, Vaz-AdvMater-2010}).  The
magnetic properties of an intrinsic bulk magnetoelectric, of which
Cr$_2$O$_3$ is a prototype, can be modulated by an external field
through the change of the magnetic cations' displacement relative to
anions \cite{Dzyaloshinskii-SPJETP-1960}.  Extrinsic magnetoelectric
couplings are typically mediated by strain: in composites of
piezomagnetic materials combined with electrostrictive materials,
external fields modulate the electric polarization, as well as the
shape of the piezoelectric. This change in turn induces strain of the
magnetic components and modifies the magnetization in the
magnetostrictive material \cite{Srinivasan-PRB-2001}.

However, although these bulk mechanisms are well understood, the
magnitude of magnetoelectric couplings in bulk materials is generally
small~\cite{Spaldin-Science-2010}, impeding their applications in
electronic devices. Moving away from bulk materials, artificial
heterostructures such as interfaces are promising candidates for
realizing or even engineering magnetoelectric couplings. Recently, a
variety of mechanisms were proposed to induce magnetoelectric coupling
at interfaces.  At ferromagnet/ferroelectric interfaces, the
interfacial bond length can be altered by the presence of
ferroelectric polarization, for example in Fe/BaTiO$_3$
\cite{Duan-PRL-2006}, Co$_2$MnSi/BaTiO$_3$ \cite{Yamauchi-APL-2007}
and Fe$_3$O$_4$/BaTiO$_3$ \cite{Niranjan-PRB-2008}.  Another mechanism
is to apply an external field and induce magnetization mediated by
free screening carriers accumulated at ferromagnetic/dielectric
interfaces, for example SrRuO$_3$/SrTiO$_3$
\cite{Rondinelli-NatNanotech-2008}.  A more complex but intriguing
mechanism was recently described theoretically and experimentally
\cite{Burton-PRB-2009,Vaz-PRL-2010} at ferroelectric/conducting
magnetic manganite interfaces.  Our work focuses on this last class of systems, 
because the coupling between ferroelectric polarization and 
magnetization not only is of great importance in fundemental sciences, 
but also finds very promising applications in memory devices
~\cite{Garcia-Nature-2009, Bristowe-PRB-2012}. 
 
At a ferroelectric/manganite interface, the presence of the
ferroelectric polarization causes screening charges to appear at the
interface due to accumulation or depletion of carriers in the
interfacial region.  Much like SrRuO$_3$/SrTiO$_3$ interfaces, the
magnetization of the interfacial atoms can be enhanced due to the
modification of carrier density around the interface because the
magnetic moment of the atoms depends on the doping level (provided
that the manganite is in the ferromagnetic phase).  What is different
in the ferroelectric/manganite system is that the accumulation of
carriers not only changes the magnetic moment but can lead to an
interfacial ferromagnetic-to-antiferromagnetic transition which
reverses the directions of the moments and thus to a much larger
magnetoelectric coupling. The work of Ref.~\cite{Burton-PRB-2009}
studied a representative heterostructure:
BaTiO$_3$/La$_{1-x}$Ba$_x$MnO$_3$ with $x=0.5$.  For bulk \lsm,
$x=0.5$ is at the critical doping level separating ferromagnetic and
antiferromagnetic phases, so the system is highly susceptible to
magnetic changes with small changes of doping.  What was found is that
when the ferroelectric polarization is flipped, the magnetic moment of
the Mn atoms in the second unit cell away from the interface is
reversed.  We note that the result is somewhat unintuitive as the
carrier doping density is highest in the first layer at the interface
which is most susceptible to change of magnetic phase.  Experiments on
the Pb(Ti$_{0.8}$Zr$_{0.2}$)O$_3$/\lsm~ are performed for $x=0.2$
~\cite{Vaz-PRL-2010} which is quite far from the boundary: {\it a
  priori} it is not clear whether enough screening charges can
accumulate to drive the system over the magnetic phase transition.
The experiments find a large magneto-electric coupling which is
interpreted to originate from a spin-flip in the first unit cell of
the manganite closest to the interface~\cite{Vaz-PRL-2010}.

In this work, we comprehensively study this proposed magnetoelectric
phenomenon at the ferroelectric/manganite interface using
first-principles calculations. We choose
PbTiO$_3$/La$_{1-x}$Sr$_x$MnO$_3$ as our prototype for computational
convenience. Although the experiments use Pb(Ti$_{0.8}$Zr$_{0.2}$)O$_3$
as the ferroelectric \cite{Vaz-PRL-2010} in order to reduce leakage
currents, the key role of the ferroelectric is to induce screening
charges due to its surface polarization, and we believe the
compositional details are secondary. 
For the \lsm~thin film, we choose the
nominal hole doping to be $x=0.2$ (unless otherwise specified) in
order to directly compare to experiment \cite{Vaz-PRL-2010}.

The original contributions of this paper are the following. (i) We are
able to show that even though the manganite film has a low nominal
doping of $x=0.2$, the ferroelectric field effects can sufficiently
change the carrier density to induce a magnetic transition at the
interface. However, the precise ground-state magnetic structure of
the interface (e.g., no spin-flip, spin-flip in first Mn layer, 
spin-flip in second Mn layer) depends sensitively on the choice of exchange
correlation functionals and specifically the value of the Hubbard $U$
parameter in the DFT+$U$ approach.  (ii) We clarify the physical reasons
of the sensitivity based on a simple Ising-like nearest-neighbor model
using bulk-derived parameters that well describes the computed
dependences. (iii) We describe an unexpected purely interfacial effect
that significantly modifies the energies of magnetic states at the
interface: the ferroelectric polarization propagates into the first
few manganite layers and the resulting rumplings of atomic layers are
responsible for the modifications. This degree of freedom is not
present in the bulk and potentially represents a further degree of
freedom that can be exploited to modify and engineer material properties
at oxide interfaces. (iv) We show that various approaches to find an
appropriate $U$ produce significantly different $U$ values, some of
which do poorly when used to compute results that compare to bulk
properties of manganites; empiricism in the choice of $U$ is an
unfortunate necessity for manganites using state-of-the-art {\it ab
  initio} methods. (v) In the process of this work, we 
develop an alternative method for counting electrons on the Mn atoms
in \lsm~thin films that is directly based on the electron density 
instead of the standard and widely used
method based on projecting onto L\"{o}wdin or atomic-like
orbitals~\cite{Lowdin-JChem-1950}. 
With this method, we can quantify the carrier distribution as a function of
ferroelectric polarization and calculate the layer-resolved effective 
hole doping. The method is generally applicable to
half-metallic oxide films.

The remainder of the paper is organized as follows. 
We discuss computational details in 
Section~\ref{computation}. 
We first study the magnetic phase transition of bulk \lsm~in 
Section~\ref{phase}. A discussion of Hubbard $U$ for bulk 
\lsm~is presented in Section~\ref{choice}.  
The charge modulation at the interface is studied in Section~\ref{charge} and 
Section~\ref{magnetization} is devoted to the discussion of spin 
modulation at the interface. We conclude in Section~\ref{conclusion}.
A number of appendices contain further technical details.
 
\section{Computational details}
\label{computation}

Our calculations are performed using density functional theory within 
the \textit{ab initio} supercell plane-wave approach~\cite{Payne-RMP-1992},
with the code PWscf in the Quantum-ESPRESSO package
\footnote{See http://www.quantum-espresso.org}. 
We employ ultrasoft pseudopotentials~\cite{Vanderbilt-PRB-1990}. The 
semicore states and reference configuration of each element are shown
in Table \ref{tab:psp}.
\begin{table}
\caption{\label{tab:psp} The semicore states and reference configurations 
of our pseudopotentials. The cut-off radii are in units of Bohr.}
\begin{center}
\begin{tabular}{c|c|c|c|c}
\hline
\hline
Atom &  Reference valence states  & $r^s_c$ &  $r^p_c$ & $r^d_c$ \\ 
\hline 
Pb      &  $5d^{10}6s^{2}6p^{2}$             & 2.5   &  2.5   & 2.3  \\
\hline
Ti       &   $3s^{2}3p^{6}3d^{1}4s^{2}$   & 1.8   &  1.8   & 1.8  \\
\hline
Sr      &   $4s^{2}4p^{6}5s^{2}$               & 2.0   &  1.8   & 2.0  \\
\hline
La      &   $5s^{2}5p^{6}5d^{1}6s^{1.5}6p^{0.5}$   &  2.2   &  2.0  & 2.2  \\
\hline
Mn     &   $3s^{2}3p^{6}3d^{5}4s^{2}$   & 2.0    & 2.0   & 2.0 \\
\hline
Pt       &   $5d^{9}6s^{1}6p^{0}$              & 1.0   & 1.0    & 1.2 \\
\hline
O        &   $2s^{2}2p^{4}$                         & 1.3   & 1.3    & --   \\
\hline\hline
\end{tabular}
\end{center}
\end{table}
We use the local spin density approximation (LSDA)~\cite{Kohn-PR-1965} for 
the exchange correlation functional as well as the Hubbard 
$U$ correction method (LSDA+$U$)~\cite{Anisimov-Condense-1997} to account 
for some of the strong 
electronic correlations on the localized $d$ orbitals of Mn atoms.  
The plane wave basis energy cutoff and charge cutoff are
35 Ry and 280 Ry, respectively. We use a Gaussian smearing width
of 5 mRy when sampling the Brillouin zone. For bulk \lsm, the $k$-grid 
sampling of the Brillouin zone is $20\times20\times 20$ per formula unit. 
For interface calculations, the $k$-grid sampling is 
$20\times20\times2$ 
where the $z$-axis is orthogonal to the interface. 
For variable cell relaxations, 
the convergence threshold for pressure is 0.5 Kbar. 
For atom relaxations, the convergence threshold for every force component 
is 26 meV/\AA. We have checked the convergence in total energies and 
structural parameters
by further increasing the $k$-point sampling 
and reducing the stress and force threshold, and observe no significant 
differences in key physical observables.

The $A$-site La$_{1-x}$Sr$_{x}$ alloying is treated by the virtual 
crystal approximation \cite{Nordheim-AnnPhys-1931, Vanderbilt-PRB-2000}.
Appendix \ref{virtualxtal} describes tests on the accuracy of the virtual
crystal approximation for our system: the results are highly satisfactory
and consistent with earlier observations~\cite{Fang-PRL-2000}

\section{Bulk manganites}
\label{bulk}

\subsection{Phase transition of magnetic ordering}
\label{phase}

\begin{figure}[t]
\includegraphics[angle=-90,width=13cm]{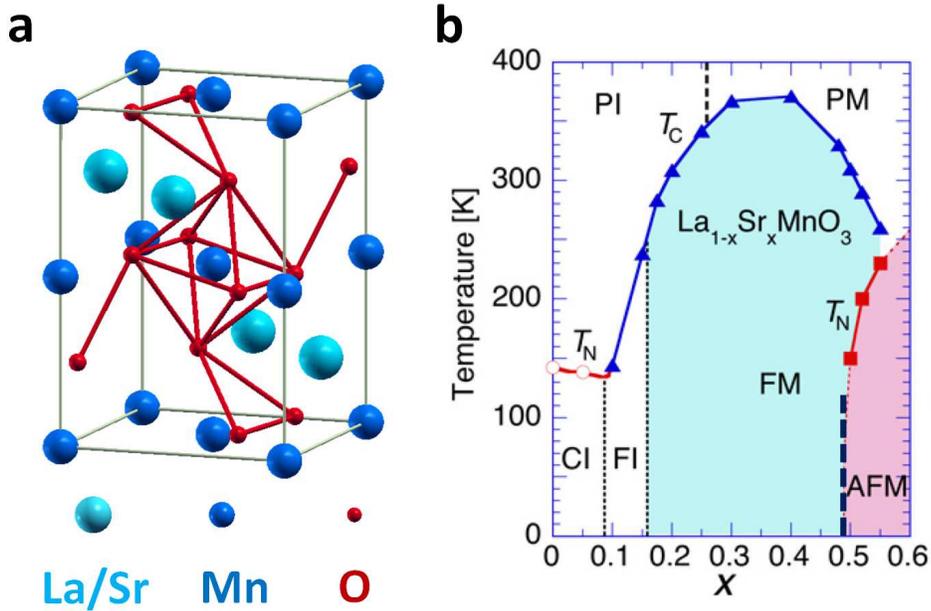}
\caption{\label{fig:phase_diagram} (Color online) \textbf{a)} 20-atom unit cell 
of \textit{Pnma} La$_{1-x}$Sr$_x$MnO$_3$. \textbf{b)} Experimental 
phase diagram of La$_{1-x}$Sr$_x$MnO$_3$. `CI' means canted insulating. 
`FI' means ferromagnetic insulating. `FM' means ferromagnetic metallic.
`AFM' means antiferromagnetic metallic. `PI' means paramagnetic insulating.
`PM' means paramagnetic metallic. The phase boundary of 
the ferromagnetic to $A$-type antiferromagnetic phase transition at low 
temperatures is highlighted by the bold dashed line. 
The panel \textbf{b)} 
is reproduced with permission from Ref.~\cite{Tokura-RepProgPhys-2006}. 
Copyright 2006 Institute of Physics Publishing.}
\end{figure}

Since, as we will show, the magnetic properties of 
PbTiO$_3$/La$_{1-x}$Sr$_x$MnO$_3$ 
interface can be understood qualitatively in terms of bulk \lsm, we start 
the discussion  with the phase diagram of bulk \lsm \phantom{} 
under different conditions. The parent compound of \lsm 
\phantom{ } is 
LaMnO$_3$ which is an $A$-type antiferromagnetic Mott 
insulator~\cite{Nohara-PRB-2006}. 
Bulk LaMnO$_3$ has strong Jahn-Teller and 
GdFeO$_3$ distortions with $Pnma$ symmetry \cite{Trimarchi-PRB-2005} 
and its primitive cell is of size 
$c(2\times2)\times 2$ in 
units of the cubic perovskite. The smallest unit cell of LaMnO$_3$ 
has four formula units (20 atoms), as is illustrated in 
Fig.~\ref{fig:phase_diagram}a. 
Chemically doping LaMnO$_3$ with Sr induces
holes on the Mn $d$-orbitals, leading to conduction and various magnetic 
orderings. Fig.~\ref{fig:phase_diagram}b shows the experimental phase 
diagram of bulk La$_{1-x}$Sr$_{x}$MnO$_3$. A ferromagnetic to $A$-type 
antiferromagnetic phase transition occurs around $x=$0.5 doping, which 
is highlighted by the bold dashed line in Fig.~\ref{fig:phase_diagram}b. 
For a random alloy distribution, we assume that bulk La$_{1-x}$Sr$_x$MnO$_3$
has the same symmetry as LaMnO$_3$ (\textit{Pnma}).
In DFT simulations, we replace La with the fictitious atom 
La$_{1-x}$Sr$_x$ in the virtual crystal approximation and 
calculate the energy difference between ferromagnetic ordering ($F$) and 
$A$-type antiferromagnetic ordering ($A$) as a function of doping $x$.

\begin{equation}
\label{equ:deltaE}
\Delta E = E(A) - E(F)
\end{equation}
In addition to the doping dependence $x$, we also study the effect of strain, 
structural distortions and Hubbard $U$ on the magnetic transition of bulk \lsm.

\begin{figure}[t]
\includegraphics[angle=-90,width=13cm]{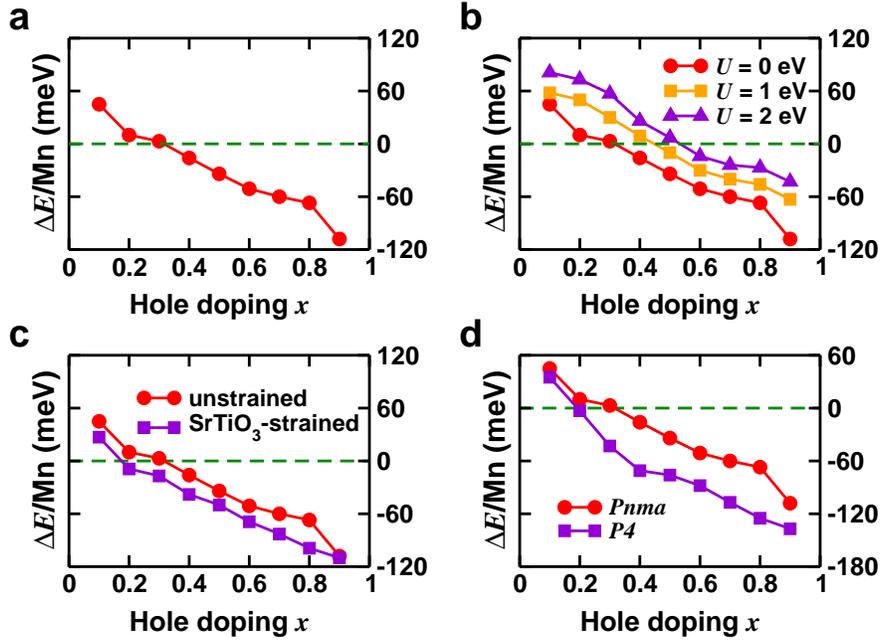}
\caption{\label{fig:bulk_lsmo} (Color online) Phase diagram of bulk $Pnma$ \lsm. $\Delta E$ is defined 
in Eq.~(\ref{equ:deltaE}). Above zero, the ground state is ferromagnetic and 
below zero it is $A$-type antiferromagnetic.
\textbf{a)} Hole doping  
dependence within LSDA. \textbf{b)} Hubbard $U$ dependence. 
\textbf{c)} Strain dependence. 
\textbf{d)} Structural distortion dependence.}
\end{figure}

\subsubsection{Doping dependence}
\label{doping}

Fig. \ref{fig:bulk_lsmo}a shows a representative bulk phase diagram of
\lsm \phantom{ } as a function of doping $x$. The calculation is
performed in the LSDA (i.e.  Hubbard $U$=0). DFT-LSDA reproduces the
ferromagnetic-to-antiferromagnetic phase transition that is observed
in experiment. The doping dependence can be understood as follows: the
ferromagnetic ordering is stablized by the double exchange mechanism
which relies on the hopping of the lone $e_g$ electrons among the
Mn$^{3+}$ ions~\cite{Zener-PR-1951}. With the increasing hole 
doping $x$, the itinerant
electrons (i.e. Mn$^{3+}$ ions) are drained and the hopping processes
are suppressed. Therefore the double exchange mechanism becomes less
operative and the ferromagnetic ordering gets more unstable as $x$ is
increased. We need to point out that in the experimental phase
diagram for $x<0.1$, \lsm \phantom{} is a spin-canted insulator (a
magnetic structure close to $A$-type antiferromagnetic ordering but 
the magnetic moment is not exactly cancelled due to weak spin-canting)
\cite{Tokura-RepProgPhys-2006}. As $x$ increases to 0.2, there is an
insulating-to-metallic transition and the appearance of the ferromagnetic 
ordering which is not reproduced in the
DFT-LSDA calculation, since the LSDA ground state is metallic 
in the whole doping range. Turning on the Hubbard $U$ does not 
change the metallicity of ferromagnetic \lsm. We 
argue that even though LSDA or LSDA+$U$ is not accurate enough to reproduce 
the spin-canted ground state at very low doping 
nor the 
insulating-to-metallic transition around $x\simeq 0.2$, it correctly
produces the metallic ferromagnetic to metallic antiferromagnetic phase 
transition at larger $x$, which 
is the key to understanding the spin-modified magnetic structure of manganites 
in the presence of ferroelectrics.
   
\subsubsection{Hubbard $U$ dependence}
\label{hubbard}

Fig. \ref{fig:bulk_lsmo}b shows a comparison of LSDA and LSDA+$U$ calculations 
for bulk \lsm. The ferromagnetic-to-antiferromagnetic 
phase transition is well reproduced  
in both LSDA and LSDA+$U$ calculations, but the transition point, i.e. 
the critical hole density where the ground state changes magnetic 
ordering, depends 
on the value of the Hubbard $U$. With an increasing $U$, the 
transition point moves to larger doping values
while the overall features of the transition remain unchanged. 
The Hubbard $U$ dependence originates as follows: 
antiferromagnetism is favored by the superexchange mechanism which 
involves the virtual hopping of electrons between low and high energy 
sites with the same spin~\cite{Kramers-Physica-1934}. 
A larger $U$ increases the energy splitting and 
thus the virtual hopping is suppressed. 
Therefore the superexchange mechanism 
is suppressed as $U$ increases, and the antiferromagnetic ordering 
accordingly becomes less stable, resulting in the upward shift of phase 
transition curve (favoring ferromagnetism). 
Empirically in order to correctly locate the transition 
point at the experimental value of $x\simeq 0.5$, we need a Hubbard $U$ in 
the range of 1 eV$<U<$2 eV in the LSDA+$U$ approximation (as illustrated in 
Fig. \ref{fig:bulk_lsmo}b).

\subsubsection{Strain dependence}
\label{strain}

Since the \lsm~thin film is grown coherently on an
SrTiO$_3$ substrate, we also study the phase diagram of
SrTiO$_3$-strained \lsm~and compare it with unstrained
bulk \lsm~in Fig. \ref{fig:bulk_lsmo}c. 
\lsm~in the whole doping range
is under weak tensile strain (within 1\%) when on an
SrTiO$_3$ substrate. Tensile (compressive) strain removes the
degeneracy of Mn $e_g$ orbitals and favors $d_{x^2-y^2}$
($d_{3z^2-r^2}$) orbitals due to the change of crystal field 
\cite{Sadoc-PRL-2010}. 
Based on the double exchange mechanism, 
ferromagnetism is isotropic with equal hoppings between Mn atoms along 
$x$, $y$ and $z$ directions. $A$-type antiferromagnetism is 
ferromagnetic in-plane and alternates its spin orientation layer by layer 
along the out-of-plane axis~\cite{Solovyev-PRB-2004}. 
Due to the tensile strain, the occupancy of $d_{3d^2-r^2}$ is lowered 
and the hopping between Mn atoms becomes essentially two-dimensional, 
suppressing ferromagnetism.  
Therefore with tensile strain, ferromagnetism is destablized
and the whole transition curve is shifted downwards
(favoring $A$-type antiferromagnetic ordering) as seen
in Fig. \ref{fig:bulk_lsmo}c.

\subsubsection{Structural distortions}
\label{structural}

Distortions away from cubic symmetry play a crucial role in the magnetism 
of manganites \cite{Solovyev-PRL-1996}.
Bulk \lsm \phantom{} has complicated structural distortions with
$Pnma$ symmetry (the unit cell is $c(2\times2)\times2$ with 20
atoms). However, we also theoretically study `artificial' \lsm
\phantom{} with only tetragonal distortions (the symmetry is $P4$ and the
unit cell is $1\times1\times2$). 
The main reason we consider the high symmetry phase ($P4$) and compare it 
to the low symmetry structure ($Pnma$) is computational: the $P4$ symmetry 
allows for the use of a smaller $1\times 1$ interface unit cell which 
allows for 
simulation of much thicker films and substrates. Therefore we need to 
understand 
the main differences, if any, between the two phases for what follows below. In 
addition, a comparison allows us to elucidate the role of structural 
distortions.

Fig. \ref{fig:bulk_lsmo}d shows a representative 
phase diagram versus doping for both $Pnma$ \lsm \phantom{} 
and $P4$ \lsm \phantom{} 
in the LSDA approximation. To understand these results, we begin with the fact 
that the effective hopping matrix element $t$ between 
neighbouring Mn atoms depends on the Mn-O-Mn bond angle 
\cite{Anderson-PR-1955}. 
In the $P4$ case, 
the bond angle is 180$^{\circ}$ and the hopping is maximized, while 
in the $Pnma$ case, the bond angle is smaller than 180$^{\circ}$ and 
the hopping is reduced. The double exchange mechanism depends linearly 
on this effective hopping matrix element $t$, while the superexchange 
mechanism lowers the energy of antiferromagnetism by $\propto t^2$ from  
second-order perturbation theory \cite{Anderson-PR-1955}. 
Therefore as we increase the hopping matrix element $t$, superexhange is 
more significantly enhanced than double exchange, thus favoring 
antiferromagnetism. Compared to the $Pnma$ case, the $P4$ case has a larger 
effective hopping and the transition curve is shifted to favor
antiferromagnetic ordering. This trend holds for both LSDA and LSDA+$U$ 
calculations. Therefore, phenomenologically we can map $Pnma$ \lsm \phantom{ } 
to $P4$ \lsm \phantom{} by choosing 
an appropriate Hubbard $U$. We find that in order to reproduce the 
ferromagnetic-to-antiferromagnetic transition around $x\simeq 0.5$ in the 
$P4$ case, we need 3 eV $< U <$ 4 eV. By comparison, to locate the 
correct transition point for $Pnma$ \lsm, $U$ must be in the range of 1 eV 
$< U <$ 2 eV (see Fig. \ref{fig:bulk_lsmo}b).

\subsection{Choosing Hubbard $U$}
\label{choice}

The DFT+$U$ approach is commonly used to study manganites
~\cite{Trimarchi-PRB-2005}. However, neither the choice of Hubbard $U$
value nor the method of choosing it is unanimous. Obviously, one can
choose $U$ based on purely empirical considerations that use
experimental data: for example, we showed above that when 1 eV $<U<$ 2
eV, LSDA+$U$ can correctly locate the experimental critical doping
density ($x=0.5$) separating ferromagnetic and antiferromagnetic
phases for bulk $Pnma$ manganites.  Below, we discuss two other
reasonable-seeming methods one might consider to determine $U$.  The
approaches yield very different values of $U$ that tend not to overlap
and do not do well in comparison to experiment. In our opinion,
unfortunately there is no reliable way to determine $U$ in a
theoretical {\it a priori} manner.  Our opinion is that a
single-particle approach such as DFT+$U$ will generally run into
difficulties in describing strongly correlated system such as
manganites, so that empiricism in choosing parameters is a necessary
fact of life.  Since the magnetic properties depend sensitively on the
value of $U$, in our mind a more fruitful approach is to study a wide
range of $U$ to understand the trends versus $U$ and especially {\it
  why} the trends take the form that they do instead of trying to make
specific predictions based on some particular choice of $U$.  (The $U$
dependence of bulk manganite and ferroelectric/manganite interfaces
are discussed in Sections \ref{hubbard} and \ref{magnetization},
respectively.)

\subsubsection{Bulk LaMnO$_3$}

First we may ask what $U$ value properly describes the parent 
material: bulk LaMnO$_3$.  This value then may be a reasonable guess
for the doped manganites.  
Taking into account the structural distortions by using
a $c(2\times2)\times2$ unit cell \cite{Trimarchi-PRB-2005} and by relaxing all degrees of freedom, 
we calculate the total energies of different magnetic orderings and 
find their energy sequence as a function of $U$. The result is shown in Table
\ref{tab:lmo}. In particular, we explicitly list $\Delta E$, defined 
by Eq.~(\ref{equ:deltaE}) in the table. We can see that within a
wide range of $U$, the ground state is not the experimentally observed
$A$-type antiferromagnet, nor is there any tendency that
ferromagnetism could yield to antiferromagnetism in the large
$U$ limit. However, reproducing the insulating properties of the
$A$-type antiferromagnetic phase requires $U\geq 4$ eV. 

We note that one can perform self-consistent calculations on bulk LaMnO$_3$
using the experimental lattice parameters and atomic coordinates.  It is
possible to stabilize an insulating $A$-type ground-state for $U\le1$ eV,
as shown in Table~\ref{tab:lmoU}.  For a  comprehensive study
of bulk LaMnO$_3$ studied with a variety of exchange correlation
functionals and basis sets, please refer to \cite{Hashimoto-PRB-2010} and 
references therein.  Unfortunately, the reproduction for the correct
ground state when using experimental structures is not of great value for 
our study: we have a non bulk-like interfacial system where the in-plane
lattice constants are fixed via epitaxy to a substrate and all remaining
degrees of freedom must be relaxed, so we must return to Table~\ref{tab:lmo}.
It would seem the best choice is either $U=0$ (which stabilizes the
incorrect ground-state by the least energy) 
or $U>4$ (which makes the $A$-type phase
insulating).  As shown above, neither choice is satisfactory in reproducing the 
experimental $x\simeq0.5$ phase boundary for the doped manganites.

\begin{table}[h]
\caption{\label{tab:lmo} LSDA+$U$ study of bulk LaMnO$_3$. The italic
  $M$ means metallic and the itallic $I$ means insulating. $F$ refers to 
  ferromagnetic ordering. $A$, $C$ and $G$ refer to $A$-type, $C$-type 
  and $G$-type antiferromagnetic ordering, 
  respectively~\cite{Solovyev-PRB-2004}. $\Delta E$
  is the energy difference between the ferromagnetic ordering and
  $A$-type antiferromagentic ordering per Mn atom, defined by
  Eq.~(\ref{equ:deltaE}).  The unit cell is orthorombic.  The
  experimental value of lattice constants is: $a=5.742$ \AA, $b=7.668$
  \AA\ and $c=5.532$ \AA~\cite{Trimarchi-PRB-2005}. The calculated
  lattice constants are for $A$-type antiferromagnetic ordering
  because the experimental ground state is $A$-type
  antiferromagnetic. In the parenthesis lists the relative difference 
  between experimental and theoretical lattice constants.}
\begin{center}
\begin{tabular}{c|c|c|c|c|c}
\hline
\hline  
$U$ (eV)  &  magnetic ordering         &  $\Delta E$ (meV)   &  $a$(\AA) & $b$(\AA) & $c$(\AA)   \\  
\hline 
0     &  $F(M)<A(M)<C(M)<G(M)$   &  40  & 5.402 (-5.9\%) & 7.468 (-2.6\%) & 5.458 (-1.3\%) \\ 
\hline
2     &  $F(M)<A(M)<C(M)<G(I)$   &  61  & 5.567 (-3.0\%) & 7.560 (-1.4\%) & 5.435 (-1.8\%) \\ 
\hline
4     &  $F(M)<A(I)<C(I)<G(I)$   &  65  & 5.644 (-1.7\%) & 7.584 (-1.1\%) & 5.448 (-1.5\%) \\
\hline
6     &  $F(M)<A(I)<C(I)<G(I)$   &  76  & 5.699 (-0.8\%) & 7.624 (-0.6\%) & 5.465 (-1.2\%) \\
\hline
8     &  $F(M)<A(I)<C(I)<G(I)$   &  98  & 5.743 (0.02\%) & 7.694 (0.3\%)  & 5.482 (-0.9\%) \\
\hline
\hline
\end{tabular}
\end{center}
\end{table}

\begin{table}[h]
\caption{\label{tab:lmoU} LSDA+$U$ study of bulk LaMnO$_3$ using experimental
coordinates and lattice constants. The italic $M$ means metallic and 
the itallic $I$ means insulating.  $\Delta E$ is the energy difference 
between the ferromagnetic ordering and $A$-type antiferromagentic ordering 
per Mn atom, defined by Eq.~(\ref{equ:deltaE}). }
\begin{center}
\begin{tabular}{c|c|c}
\hline
\hline  
$U$ (eV)  &  magnetic ordering         &  $\Delta E$ (meV)   \\  
\hline 
0     &  $A(I)<F(M)$   &  -15  \\ 
\hline
1     &  $A(I)<F(M)$   &   -4  \\ 
\hline
2     &  $F(M)<A(I)$   &    4  \\
\hline
3     &  $F(M)<A(I)$   &   10  \\
\hline
4     &  $F(M)<A(I)$   &   17  \\
\hline
\hline
\end{tabular}
\end{center}
\end{table}

\subsubsection{Linear response approach of self-consistent $U$}

Second, we may ask for a purely {\it ab initio} approach that delivers
a value of $U$ appropriate for the system within the framework of
DFT+$U$ itself.  This is the linear response approach of
Refs.~\cite{Cococcioni-PRB-2005, Kulik-PRL-2006}.  We focus on $P4$
\lsm~($x=0.2$) as an example. We run a series of linear response calculations
\cite{Cococcioni-PRB-2005} on $2\times 2\times 2$ unit cells of $P4$
\lsm. The ground state is calculated using LSDA+$U$ with a range of $0
< U_{in} < 5$ eV. For each value of $U_{in}$, we use the extrapolation
scheme in Ref. \cite{Cococcioni-PRB-2005} to get the converged value
of $U_{out}$.  Then we collect all the converged $U_{out}$ as a
function of $U_{in}$ and extract out $U_{scf}$ \cite{Kulik-PRL-2006}
from the linear region.  Our final value is $U_{scf}=5.8$ eV.  As
discussed above, to reproduce the experimental $x\simeq0.5$ boundary
for $P4$ \lsm, we require 3 eV$<U<4$ eV.  The self-consistent $U$ is
significantly higher.

\section{Ferroelectric/manganite interfaces}
\label{interface}

\subsection{Methodology}

\begin{figure}[t]
\includegraphics[angle=-90,width=15cm]{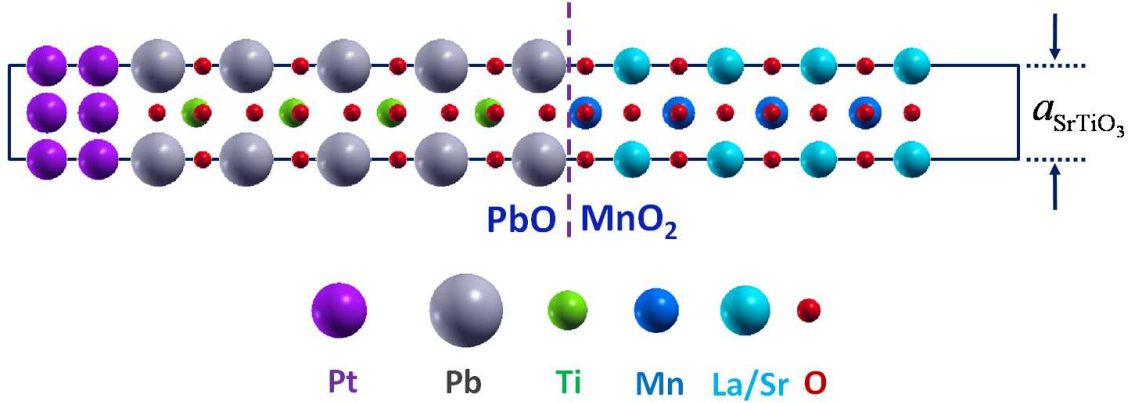}
\caption{\label{fig:supercell} (Color online) Illustration of the computational supercell. 
The dashed line highlights the PbO/MnO$_2$ interface. The whole structure is 
coherently strained to the lattice constant of SrTiO$_3$. 
Vacuum $\simeq$ 20~\AA~thick is introduced to separate 
periodic copies. An $X$O ($X$=La/Sr) atomic layer faces the vacuum.}
\end{figure}

Our computational supercell for interface calculations is schematically 
illustrated in 
Fig.~\ref{fig:supercell}. The $x$ and $y$ directions of the 
simulation cell are subject to periodic boundary conditions 
and their lengths are fixed
to our computed theoretical lattice constant of SrTiO$_3$ $a=3.85$~\AA~(1.5\%
smaller than the experimental value), because in experiments \lsm~is
epitaxially grown on a SrTiO$_3$ substrate~\cite{Vaz-PRL-2010}.  
In order to directly compare to the experiment, all the results shown
below are from calculations with the nominal doping level chosen as
$x=0.2$ (unless otherwise specified).  In addition to PbTiO$_3$ and
\lsm, we also include the electrode Pt to provide an electron
reservoir and $\simeq$ 20~\AA~vacuum to separate periodic copies of
the slabs.  We strain the in-plane lattice constant of the entire slab
structure to that of bulk SrTiO$_3$ to impose the epitaxial strain
from the substrate.  In the simulation cell (Fig.~\ref{fig:supercell})
and in most of our calculations, we do not include a SrTiO$_3$
substrate explicitly in order to keep the computations from becoming
unwieldy in scale. However, in Appendix~\ref{sto} we present a few
calculations that do include the SrTiO$_3$ substrate explicitly, and
it is shown that the interfacial structural and magnetic properties
between PbTiO$_3$/\lsm~ are well converged when the \lsm~film is 4
unit cells or thicker.  In addition to reducing the computational
burden, the absence of a SrTiO$_3$ substrate creates a manganite
surface that allows us to apply a hole counting method which can much
more accurately calculate the hole spatial distribution than the use
of L\"{o}wdin orbitals~\cite{Lowdin-JChem-1950} (see
Appendix~\ref{count} for details). Since the SrTiO$_3$ substrate is
typically TiO$_2$-terminated and the manganites are in principle
deposited stoichiometrically and epitaxially on the SrTiO$_3$, the
resulting ferroelectric/manganite interface is taken to be
PbO/MnO$_2$.

Using the Berry phase method \cite{King-Smith-PRB-1993}, we find that
SrTiO$_3$-strained PbTiO$_3$ has bulk polarization 0.74 C/m$^2$.  For
the two different directions of ferroelectric polarization, we define
two distinct states: the accumulation state in which extra holes are
induced into the interfacial \lsm~and depletion state in which extra
electrons are induced into the interfacial \lsm~(i.e., holes are
driven out).  One unit cell of PbTiO$_3$ in the interior is fixed to
the bulk ferroelectric PbTiO$_3$ positions, an choice that simulates
the behavior of a thick PbTiO$_3$ film. All remaining atomic
coordinates in the slab are relaxed. We need to mention that our choice of 
boundary condition on ferroelectrics is consistent with the 
experiment~\cite{Vaz-PRL-2010} 
in which a thick film of 250 nm Pb(Zr$_{0.2}$Ti$_{0.8}$)O$_3$ is deposited on 
La$_{1-x}$Sr$_x$MnO$_3$. However, the boundary 
condition on the ferroelectric could be different, depending on the experiments 
to be studied. For example, 
in Ref.~\cite{Bristowe-PRB-2012}, 
three unit cells of BaTiO$_3$ 
adjacent to a La$_{0.7}$Sr$_{0.3}$MnO$_3$ film are fully relaxed 
without the presence of Pt electron reservior, 
in order to simulate the ultra-thin ferroelectrics used in other 
experiments~\cite{Garcia-Nature-2009}. Interestingly, the results 
of magnetoelectric coupling in Ref.~\cite{Bristowe-PRB-2012} are 
consistent with ours, described below.   

\begin{figure}[t!]
\includegraphics[angle=-90,width=12cm]{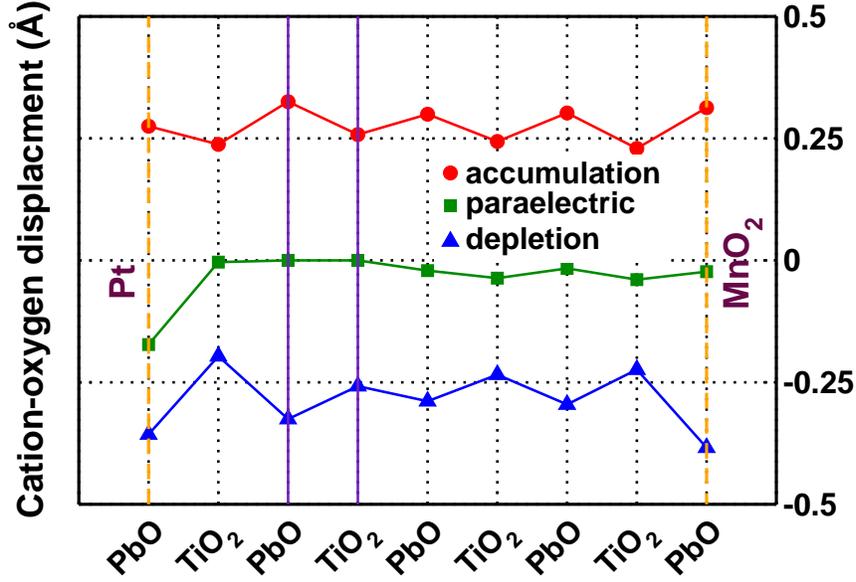}
\caption{\label{fig:polar_dis} (Color online) 
Cation-oxygen displacements along $z$-axis 
for the PbTiO$_3$ film inside the Pt/PbTiO$_3$/La$_{1-x}$Sr$_x$MnO$_3$ 
heterostructures. 
The two purple solid lines highlights the one unit cell of PbTiO$_3$ 
that is fixed to bulk positions. The two orange dashed lines show 
the interfaces: left is the one facing Pt electrodes and right is the one 
that faces \lsm.}
\end{figure}

In Fig.~\ref{fig:polar_dis}, we show the 
cation-oxygen $z$-axis displacements of a representative PbTiO$_3$
thin film within the slab structure. The single fixed unit cell is
highlighted by the two solid purple lines. The two interfaces (one
faces \lsm~and the other faces Pt electrodes) are shown by the orange
dashed lines. The sign of the displacements indicates polarization
directions.  We can see that in the relaxed PbTiO$_3$, there is no
reversal of ferroelectric polarization and the magnitude of
polarization is homogeneous. In addition, we also calculate an
artificial state in which one unit cell of PbTiO$_3$ is fixed to be
paraelectric (i.e., zero cation-oxygen rumpling in the (100) atomic
plane of the fixed unit cell). 

Finally, we mention that most of the results presented below are
calculated for an in-plane $c(2\times2)$ unit cell which is compatible
with the structural distortions found in bulk $Pnma$ \lsm. Such
calculations are referred to as $c(2\times2)$ \lsm~interface
calculations. In order to converge the hole distribution versus \lsm~
thickness without inordinately increasing the computational burden, we
increase the thickness of manganites by reducing the in-plane cell to
$1\times 1$. Those calculations are referred to as $1\times 1$
\lsm~interface calculations.

\subsection{Charge modulation}
\label{charge}

We first study the effect of charge modulation from switching the
ferroelectric polarization of PbTiO$_3$. In the presence of ferroelectric 
PbTiO$_3$, the charge density of \lsm \phantom{} at the interface 
differs from its bulk value because the polarization of PbTiO$_3$ terminates at the interface and
results in the surface charge (the surface charge density is
$\sigma=\textbf{P}\cdot\textbf{n}=P_z$). Since \lsm \phantom{} is metallic, 
this surface charge induces
screening charge in the \lsm \phantom{} equal in magnitude 
but opposite in sign to the surface charge. 
When the PbTiO$_3$ switches its polarization, the surface
charge changes sign and so does the screening charge. Therefore
a net change of charge density ($\Delta \sigma = 2P_z$) is 
induced in the \lsm~thin film. 

\begin{figure}[t]
\includegraphics[angle=-90,width=15cm]{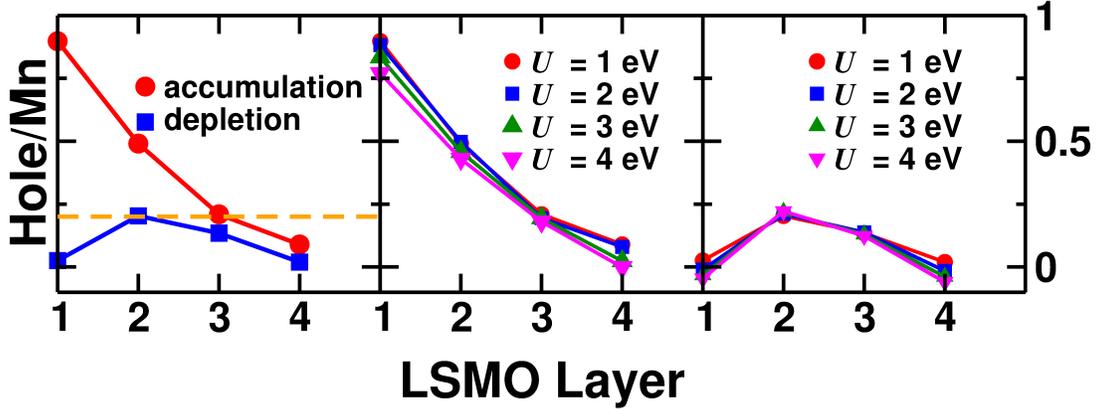}
\caption{\label{fig:hole_distribution_U} (Color online) 
Left panel: hole distributions 
of the accumulation and depletion states of $c(2\times2)$ 
\lsm~($U$ = 1 eV). Middle panel: Hubbard $U$ dependence of hole 
distribution in the accumulation state. Right panel: Hubbard $U$ dependence 
of hole distribution in the depletion state.}
\end{figure}

The left panel of Fig. \ref{fig:hole_distribution_U} shows
the hole distribution of a 4 unit cells thick \lsm~film on PbTiO$_3$. 
The method of counting holes is described in the Appendix \ref{count}.  
The nominal doping is $x=0.2$, highlighted by the dashed line. This 
calculation is performed on $c(2\times 2)$ \lsm~with LSDA+$U$ 
($U = 1$ eV). The PbO/MnO$_2$ interface is at layer 1, and layer 4 is the 
artificial surface. As expected, the hole distribution
accumulates (depletes) at the interface when the surface charge is
negative (positive).
 
The middle and right panel of Fig. \ref{fig:hole_distribution_U} show 
the Hubbard $U$ dependence of spatial hole distribution for the 
accumulation and depletion states, respectively. The 
calculation is performed on the same structure as in the left panel of Fig. 
\ref{fig:hole_distribution_U}.
Since the induced holes (or electrons) extend into the \lsm~
within the screening length (which does not strongly depend on the 
correlation), it is not surprising that Hubbard $U$ does not 
significantly change the hole distribution.
As a good approximation, we assume that the hole distribution does not 
depend on Hubbard $U$.

\begin{figure}[t]
\includegraphics[angle=-90,width=15cm]{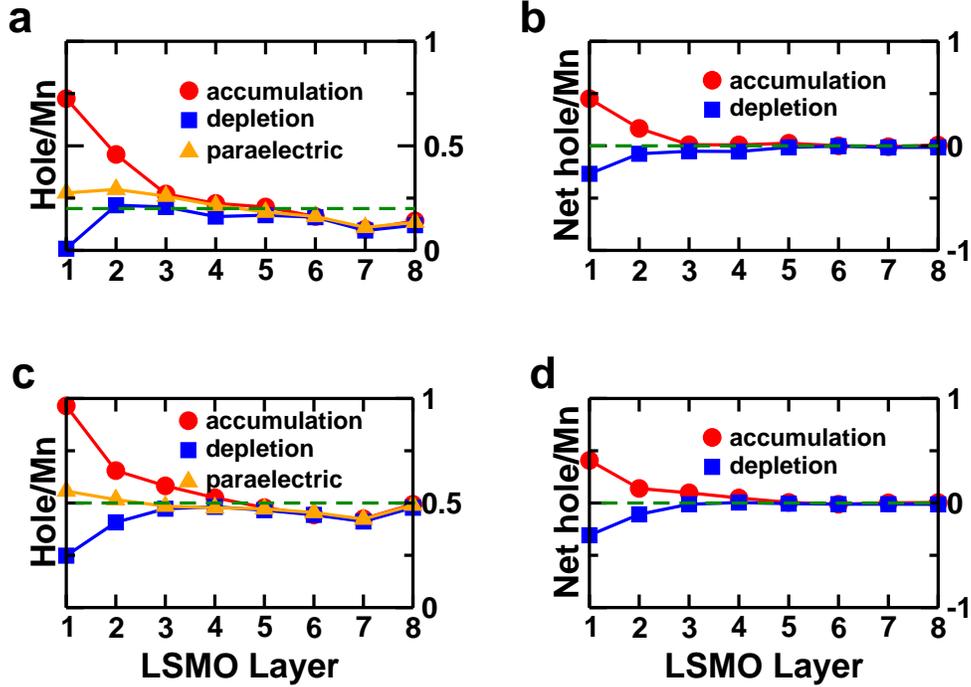}
\caption{\label{fig:hole_distribution} (Color online) \textbf{a)} 
Hole distributions of $1\times 1$ \lsm~8 
unit cells thick with $x=0.2$ ($U$ = 4 eV). In addition to accumulation 
and depletion states, a test calculation in which PbTiO$_3$ is fixed to 
be paraelectric is done and the resulting hole distribution is 
shown with orange triangles.
\textbf{b)} ``Net'' hole distributions of $1\times 1$ La$_{1-x}$Sr$_{x}$MnO$_3$ 
8 unit cells thick with $x=0.2$ ($U$ = 4 eV): the paraelectric state
background (see \textbf{a)}) is substracted from the hole distributions 
of accumulation and depletion states. 
\textbf{c)} Hole distributions of $1\times 1$ La$_{1-x}$Sr$_{x}$MnO$_3$ 8 
unit cells thick with $x=0.5$ ($U$ = 4 eV). 
\textbf{d)} ``Net'' hole distributions in $1\times 1$ 
La$_{1-x}$Sr$_{x}$MnO$_3$ 8 unit cells thick with $x=0.5$ ($U$ = 4 eV).}
\end{figure}

In order to get a spatial distribution of holes that is well converged 
with the manganite thickness, we run a calculation with 8 unit 
cells of $1\times1$ \lsm~. This calculation is performed with 
LSDA+$U$ ($U$=4 eV). The results are shown in Fig.~\ref{fig:hole_distribution}.
From Fig.~\ref{fig:hole_distribution}a, 
the screening length of the accumulation state 
is estimated to be 3 unit cells while the screening length of 
depletion state seems to be only 1 unit cell. The asymmetry is 
due to the fact that there are two factors affecting the 
hole distribution. One is the induced screening charge 
and the other is the presence of PbO/MnO$_2$ interface itself. To 
demonstrate the role of the interface, we perform a test 
calculation in which the PbTiO$_3$ thin film is forced to be paraelectric 
and find (see the orange triangle symbols in 
Fig.~\ref{fig:hole_distribution}a) that the resulting 
hole distribution is not uniform nor equal to the nominial doping 
($x =0.2$) at the interface. This non-uniform hole distribution 
can be considered as a background, owing to the chemistry of the PbO/MnO$_2$ 
interface. If we substract the hole distributions of accumulation and 
depletion states from this background, we can see that the ``net'' hole
distributions of accumulation and depletion states now become more 
symmetric, with the screening length of depletion state a little larger 
than that of accumulation state (Fig.~\ref{fig:hole_distribution}b). 
This is consistent with the Thomas-Fermi picture that depletion states have 
less carriers (holes) and therefore a larger screening length. To further 
verify this Thomas-Fermi picture, we perform the same calculation with 
a layer nominal hole doping $x=0.5$ and find very similar results (see 
Fig.~\ref{fig:hole_distribution}c and d). 

\subsection{Magnetization modulation}
\label{magnetization}

\begin{figure}[t]
\includegraphics[angle=-90,width=12cm]{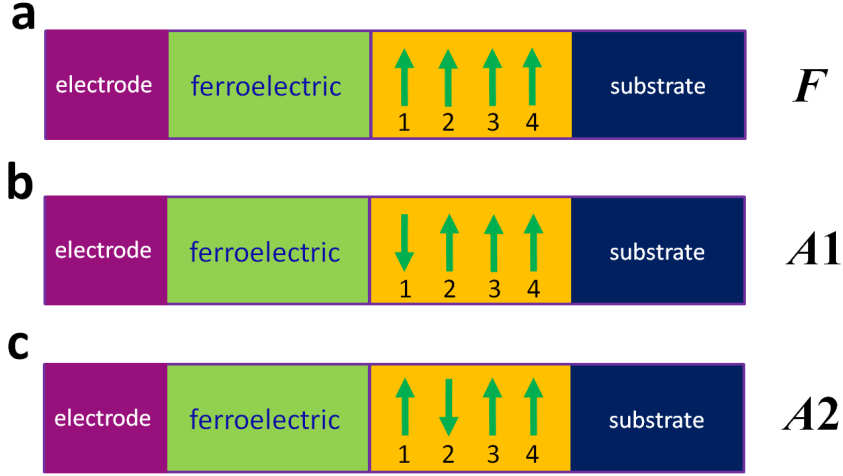}
\caption{\label{fig:mag_structure} (Color online) Illustration of different magnetic 
structures. \textbf{a)} Ferromagnetic configuration, denoted by $F$. 
\textbf{b)} Spin flips in the first unit cell, denoted by $A1$.
\textbf{c)} Spin flips in the second unit cell, denoted by $A2$.}
\end{figure}

In this section, we study in detail whether the charge modulation can
induce a spin modified configuration in the ground state.  This means
that the change of the magnetization is not simply proportional to
that of the charge density (i.e. simple filling/emptying of Mn orbitals
with fixed spin polarization) but involves a more dramatic change of
magnetic structure at the interface. The mechanism is as follows: in
the accumulation state, the local hole distribution adjacent to the
interface could be higher than $x=0.5$, the critical value for the 
ferro-to-antiferromagnetic transition. Therefore the spins at the
interfacial region could flip. However, in the depletion state,
such a local spin-flip is not expected to occur. Therefore from now on, 
we only focus on the accumulations state. 
In order to study whether this local phase transition does
occur at the interface by switching the polarization, we consider 
three relevant spin configurations ($F$, $A1$ and $A2$), illustrated in 
Fig~\ref{fig:mag_structure}.
When all the spins are ferromagnetically coupled, this configuration 
is denoted as $F$ (Fig.~\ref{fig:mag_structure}a). If the 
spin is flipped in the first unit cell of manganite from the interface, 
this configuration is denoted by $A1$ (Fig.~\ref{fig:mag_structure}b). 
Finally, if the spin is flipped in the second unit cell of manganite, 
then we denote it by $A2$ (Fig.~\ref{fig:mag_structure}c). We address 
three important and related questions below: 
i) whether the ground-state magnetic 
structure depends on $U$? ii) given a reasonable $U$, whether the 
manganite nominal doping $x$ could change the final magnetic structure? 
iii) how the structural distortions at the interface may affect the 
magnetic structures? 

\subsubsection{Hubbard $U$ dependence}
We obtain the total energies of these three spin 
configurations with a range of Hubbard $U$ and collect all the results in 
Fig. \ref{fig:energy_diff}.
We use the following definitions of energy differences:

\begin{figure}[h]
\includegraphics[angle=-90,width=12cm]{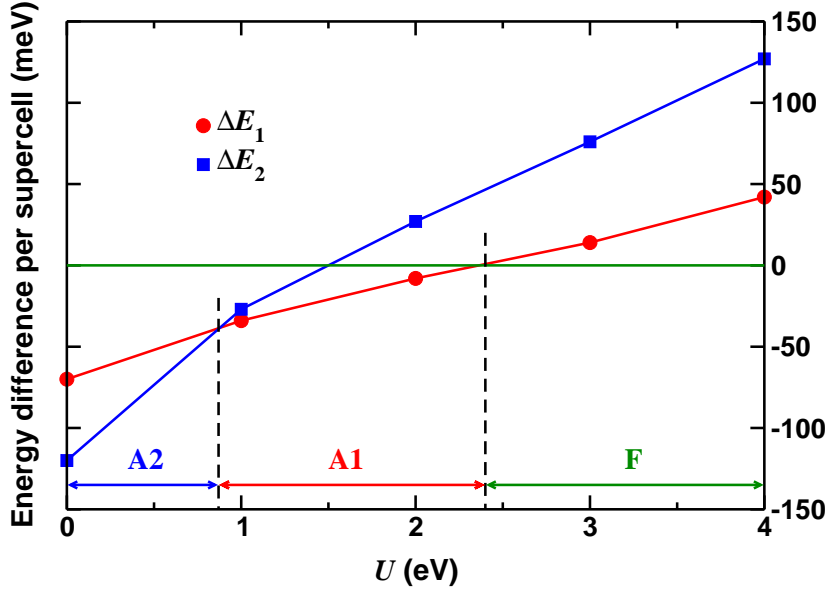}
\caption{\label{fig:energy_diff} (Color online) Energy sequence dependence on Hubbard $U$.
$\Delta E_1 = E(A1) - E(F)$ and $\Delta E_2 = E(A2) - E(F)$. The labels 
($F$, $A1$ and $A2$) show the ground state in different regions of 
Hubbard $U$. The boundary between $F$ and $A1$ is 2.4 eV and that between $A1$ 
and $A2$ is 0.9 eV.}
\end{figure}

\begin{equation} 
\label{eq-1}\Delta E_1 = E(A1)-E(F)
\end{equation}

\begin{equation} 
\label{eq-2}\Delta E_2 = E(A2)-E(F)
\end{equation}
From Fig. \ref{fig:energy_diff} we can see that the magnetic structure of
the ground state evolves with Hubbard $U$. When $U$ is
small ($U <$ 0.9 eV), the ground state has the magnetic structure of
$A2$. With $U$ increasing (0.9 eV $< U <$ 2.4 eV), the ground
state evolves into $A1$. When $U$ gets even larger ($U >$
 2.4 eV), we have $F$ as the ground state. Therefore any prediction of the
magnetic properties of the interface depends greatly on the choice of $U$.  
Before we pick a reasonable value of $U$, 
we need to understand why the magnetic
structure is so sensitive to the Hubbard $U$. The answer is that 
$U$ changes the bulk phase transition point so
that for the same hole distribution, the preferred local magnetic phase 
also changes. We can see from Fig. \ref{fig:energy_diff} that increasing 
the Hubbard $U$ drives the local phase at the interface from 
antiferromagnetic to ferromagnetic, which is consistent with 
the $U$ dependence in the bulk (see Fig. \ref{fig:bulk_lsmo}b). 
In order to more 
quantitatively describe the energy sequence, 
we construct an Ising-like model which is based on the interaction between 
nearest neighbor Mn magnetic moments~\footnote{The original Ising 
model with Hamiltonian of the form $\sum_{ij}J_{ij}S_iS_j$ can not be justified 
rigorously from first principles here because our system 
is metallic and there are no isolated spins. Instead, we use layer-resolved 
magnetization $m_i$ as a basis variable to create an effective model for our 
system which has the form of an Ising Hamiltonian.}:

\begin{equation}
\label{eq-3} E=-\sum_{<ij>}J_{ij}m_im_j
\end{equation}
where $<ij>$ range is over all nearest neighbors and $m_i$ is the magnetization 
in each MnO$_2$ layer of the manganites. The labelling of manganite layers 
is shown in Fig.~\ref{fig:mag_structure}. 
We assume that the hole spatial 
distribution does not sensitively depend on magnetic structures
\footnote{We used a L\"{o}wdin orbital analysis and found that the change in 
the hole spatial distribution between different magnetic structures is 
insignificant. However, just as we argued in the appendix, the L\"{o}wdin 
method itself is not highly accurate and can only be considered 
to provide indirect evidence.}, 
and obtain:

\begin{equation}
\label{eq-4} \Delta E_1=2J_{12}|m_1m_2|
\end{equation}
 
\begin{equation}
\label{eq-5} \Delta E_2=2J_{12}|m_1m_2|+2J_{23}|m_2m_3|
\end{equation}
In order to get an energy sequence, we need to know the signs of $J_{12}$ 
and $J_{23}$. From the bulk calculations, at a given hole doping $x$ and 
assuming half-metallicity, the magnetization is related to the hole 
doping $x$ by $m=(4-x)\mu_B$ where $\mu_B$ is the Bohr magneton. The 
exchange coupling $J$ can be extracted out by: 

\begin{equation}
\label{eq-6} J=\frac{E(A)-E(F)}{2m^2}
\end{equation}
where the energy difference $\Delta E=E(A)-E(F)$ is from the bulk 
calculations, shown in 
Fig. \ref{fig:bulk_lsmo}. $J$
changes sign at the transition point. From 
Eq.~(\ref{eq-6}), $J$ is positive for ferromagnetic phase and 
negative for $A$-type antiferromagnetic phase.
At the interface, however, the hole spatial distribution 
is not uniform (see Fig. \ref{fig:hole_distribution}). 
We assume that the interface coupling $J_{ij}$ is that of bulk 
\lsm~but for a doping value that is the average of the neighboring 
layers $i$ and $j$:

\begin{equation} 
\label{eq-7} J_{ij}\simeq J_{\textrm{bulk}}\left(\frac{x_i+x_j}{2}\right)
\end{equation}
We need a final good approximation, which is verified in Fig. 
\ref{fig:hole_distribution_U}, that the hole distribution does not 
sensitively depend on Hubbard $U$.
Based on Eq.~(\ref{eq-4}-\ref{eq-7}), 
we start with a large Hubbard $U$. 
Since large $U$ favors ferromagnetism (see Fig. \ref{fig:bulk_lsmo}b),  
the bulk phase is ferromagnetic and  
both $J_{12}$ and $J_{23}$ are positive. Thus 
$0 < \Delta E_1 < \Delta E_2$ and we have the following energy sequence:
$F< A1 < A2$. We denote this by case 1.
With a decreasing $U$, the transition point is moved to 
smaller hole doping region. Noting that the hole distribution 
monotonically decays 
from the interface (see Fig. \ref{fig:hole_distribution}), 
we always have $(x_1+x_2)/2>(x_2+x_3)/2$. Hence $J_{12}$ changes 
sign earlier than $J_{23}$ as $U$ decreases. If $U$ is in such a range 
that $J_{12}$ just becomes negative but $J_{23}>0$, we have $\Delta E_1 <0$ 
and $\Delta E_2 > 0$. The energy sequence is now $A1< F < A2$, which is 
denoted by case 2. 
As $U$ further decreases, so that $J_{12}$ becomes very negative and 
$J_{23}$ remains positive but $J_{12}|m_1|+J_{23}|m_3|<0$, then we have 
$\Delta E_1< \Delta E_2 < 0$. The energy sequence becomes $A1< A2< F$. This 
is case 3. With 
$U$ further decreasing, the 
bulk phase becomes always antiferromagnetic, both $J_{12}$ and $J_{23}$ 
become negative, and we have $\Delta E_2 < \Delta E_1 < 0$. The final 
possible energy sequence is $A2 < A1 < F$, which is denoted by case 4. 
These four energy sequences 
exhaust all the possibilities and are summarized in Table 
\ref{tab:energy_seq}.
Now we compare the DFT results (see Fig. \ref{fig:energy_diff}) 
to the energy sequence 
predicted from the model (see Table \ref{tab:energy_seq}). 
As the Hubbard 
$U$ evolves from 0 to 4 eV, we find all four cases. 
For example, $U=3$ eV corresponds to $F < A1 < A2$; $U=2$ eV to $A1 < F < A2$; 
$U=1$ eV to $A1 < A2 < F$ and $U=0$ eV to $A2 < A1 < F$. The exact 
boundaries of Hubbard $U$ for each energy sequence can be found in 
Fig. \ref{fig:energy_diff}.

\begin{table}[t]
\caption{\label{tab:energy_seq} The energy sequence predicted from the simple 
model and the comparison with the DFT calculations with different Hubbard $U$.}
\begin{center}
\begin{tabular}{c|c|c|c|c}
\hline\hline
 \multicolumn{2}{c|}{model} & \multicolumn{3}{c}{DFT} \\
\hline
case & energy sequence & $U$ (eV) & $\Delta E_1$ (meV) & $\Delta E_2$ (meV) \\ 
\hline 
1  & $0 < \Delta E_1<\Delta E_2 $   & 3   &  14   &  76  \\
\hline
2  & $\Delta E_1< 0 <\Delta E_2 $   & 2   &  -8   &  27  \\
\hline
3  & $\Delta E_1< \Delta E_2 < 0$   & 1   & -34   & -27  \\
\hline
4  & $\Delta E_2< \Delta E_1 < 0$   & 0   & -70   & -120  \\
\hline\hline
\end{tabular}
\end{center}
\end{table}

Since the Hubbard $U$ changes the transition point 
and the magnetic structure of the ground state of the 
PbTiO$_3$/La$_{1-x}$Sr$_x$MnO$_3$ interface, 
we need to determine what is the reasonable value of $U$. Following 
Ref. \cite{Burton-PRB-2009}, 
we argue that because the magnetic structures sensitively depend on the 
transition point, we need to choose a range of $U$ so that the 
ferromagnetic-to-antiferromagnetic transition occurs around 
$x\simeq 0.5$. From bulk calculations, we know that 
as 1 eV $< U <$ 2 eV for $Pnma$ \lsm, 
this criterion is satisfied. On 
the other hand, when $U$ is in this range, the magnetic structure of 
the ground state is always $A1$. Therefore by switching the PbTiO$_3$ 
polarization, we do 
find a spin-modified configuration in the DFT simulation, provided that 
our choice of $U$ is reasonable. This prediction is 
consistent with the recent experiment \cite{Vaz-PRL-2010}
which observes an anomously large change in the magnetization as the 
polarization of ferroelectrics is switched and which assigns this to a 
spin-flip on the Mn atom closest to the interface.

\subsubsection{Hole dependence}

\begin{figure}[t!]
\includegraphics[angle=-90,width=12cm]{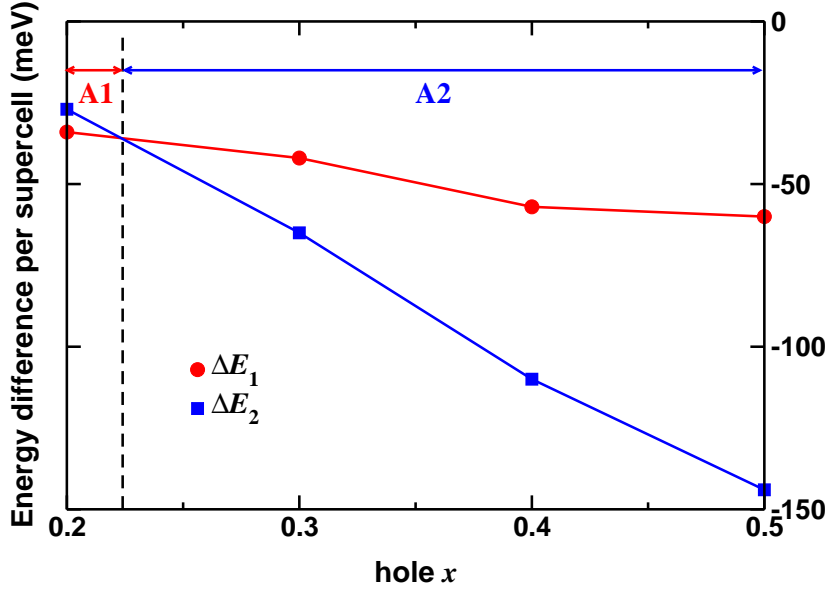}
\caption{\label{fig:energy_diff_hole} (Color online) Energy dependences of the
  interfacial magnetic states as a function of the doping $x$ of the
  La$_{1-x}$Sr$_x$MnO$_3$. $\Delta E_1 = E(A1) - E(F)$ and $\Delta E_2
  = E(A2) - E(F)$. The boundary between the phases is at $x_c\approx
  0.22$ denoted by the vertical dashed line.  $A1$ is stable to the
  left and $A2$ to the right of the boundary.  The results are based
  on LSDA+$U$ with $U=1$ eV.}
\end{figure}

The ground-state magnetic structure we found above ($A1$
configuration) is consistent with the experimental conjecture, but it
is different from the $A2$ configuration found using DFT+GGA for the
similar multiferroelectric structure BaTiO$_3$/La$_{1−x}$Ba$_x$MnO$_3$
with $x = 0.5$~\cite{Burton-PRB-2009}. We find that the reason for the
differing ground state magnetic structure is due to the doping $x$ dependence 
of the system.  Specifically, for $U = 1$ eV, we calculate the
energies of the $A1$ and $A2$ interfacial states versus doping $x$ and
present the results in Fig.~\ref{fig:energy_diff_hole}.

When the nominal doping $x$ is near the bulk magnetic transition point
$x = 0.5$, the Fig.~\ref{fig:energy_diff_hole} shows that 
the ground state should be $A2$ which
is consistent with Ref.~\cite{Burton-PRB-2009}. 
However, the actual experimental doping
$x=0.2$ is far below $x=0.5$, the ground state should be $A1$. We
calculate the phase boundary between $A1$ and $A2$ to be $x_c \approx
0.22$ for $U = 1$ eV. This transition from $A1$ to $A2$ is easily
understood in the framework of our Ising-like model. Each pair of
neighboring Mn magnetic moments could be either ferromagnetically or
antiferromagnetically coupled depending on the number of holes on the
pair of Mn atoms. When the doping $x$ is low, the ferroelectric
modulation of the hole density must drop to a low value within a few
unit cells from the interface (see Fig.~\ref{fig:hole_distribution}) 
and thus only the first two
moments $(m_1,m_2)$ will be paired antiferromagnetically, which leads
to $A1$.  However, when the doping $x$ is high, the larger values of the
hole density means that both the $(m_1,m_2$) and $(m_2,m_3)$ pairs
couple antiferromagnetically, which leads to $A2$. As a final remark,
we note that our Ising-like model predicts that the transition doping
$x_c$ between $A1$ and $A2$ should depend on Hubbard $U$: since
increasing $U$ favors ferromagnetism, larger $U$ will increase $x_c$
(i.e, require more holes for antiferromagnetism).

\subsubsection{Structural distortion dependence}

Due to the presence of ferroelectric polarization, significant 
distortions that deviate from bulk manganites are observed in the relaxed
ground state structures. Concerning each oxygen octahedron 
that encloses Mn atoms, we 
calculate $c/a$ ratio and rumplings $\delta/a$ in each MnO$_2$ layer, 
where $c$ is the distance between the two apical oxygen atoms along the 
$z$ direction, $\delta$ is the vertical displacement between Mn and O, 
and $a$ is the lattice constant of SrTiO$_3$ substrate. 
The results are summarized in Table~\ref{tab:distortions}. 
Since the spin-flipped process 
occurs at the interface in the accumulation state, we only show 
$c/a$ ratio and $\delta/a$ of the first and second manganite layers from 
the interface, and from now on the discussion is constrained to the 
accumulation state. From Table~\ref{tab:distortions}, we can see that 
in the accumulation state, there are significant polar distortions at 
the interface ($\delta/a$ is as large as 6\%).
It was shown in Ref.~\cite{Burton-PRB-2009} that the spin-flipped 
process is mainly of electronic origin rather than due to the 
polar distortions at the interface. In this section, we use detailed 
comparisons to show that though the spin-modified 
configuration is due to electronic reconstructions, polar distortions 
need to be taken into account in order to make a quantitative (instead 
of qualitative) link between the interface phase and bulk phases. 

\begin{table}[t]
\caption{\label{tab:distortions} The $c/a$ ratio of each oxygen octahedron
that encloses Mn atoms and rumplings $\delta/a$ of each MnO$_2$ layer for 
both accumulation and depletion states. $c$ is the distance between the two 
apical oxygen atoms along the $z$ direction. $\delta$ is the rumplings of 
MnO$_2$ layer, and $a$ is the lattice constant of SrTiO$_3$ substrate. Layer1 
and layer2 refer to the first and second unit cell of manganites from the 
interface.}
\begin{center}
\begin{tabular}{c|c|c|c|c|c}
\hline\hline
 \multicolumn{3}{c|}{accumulation state} & \multicolumn{3}{c}{depletion state} \\
\hline
           & layer1 & layer2 &             & layer1 & layer2\\ 
\hline 
$c/a$      & 0.97   &  0.94  &  $c/a$      &  1.05  & 0.98  \\
\hline 
$\delta/a$ & 0.06   &  0.02  &  $\delta/a$ &  0.01  & 0.001 \\
\hline\hline
\end{tabular}
\end{center}
\end{table}

Now we look at the Ising-like model Eq.~(\ref{eq-4}-\ref{eq-7}) more closely. 
The model is based on 
the assumption that the local magnetic structure can be predicted from 
bulk manganites of the same hole doping. In 
Table \ref{tab:polar_dis}, 
we list the energy difference between $F$ and $A1$ interfacial configurations 
from the interface calculations, defined by 

\begin{equation}
\label{equ:ediffI}\Delta E_{I}=E(A1)-E(F)
\end{equation} 
where $E(F)$ and $E(A1)$ are the total energies of $F$ and $A1$ 
configurations, respectively.
We also calculate the 
average hole density between the first and second layers, i.e. 
$\overline{x}=(x_1+x_2)/2$. Next, we list the \textit{bulk} 
energy difference $\Delta E_{B}$, defined as

\begin{equation}
\label{equ:ediffB}\Delta E_{B}=\frac{1}{2}\left(E_{B}(A)-E_{B}(F)\right)
\end{equation} 
where $E_B(F)$ and $E_{B}(A)$ are the total energies of 
SrTiO$_3$-strained \lsm~with ferromagnetic and $A$-type
antiferromagnetic ordering, respectively. The factor $\frac{1}{2}$ is 
included because in the bulk form wherever a Mn atom flips its spin, 
there are two Mn-Mn bonds involved owing to periodic boundary 
conditions, whereas at the interface a Mn spin flip only involves 
one Mn-Mn bond. Hence, we need a factor $\frac{1}{2}$ so that both
$\Delta E_I$ and $\Delta E_B$ describe the energy difference $per$ 
Mn-Mn bond.
The nominal hole 
doping $x$ is chosen as the same as $\overline{x}$ from the supercell 
calculations. Table~\ref{tab:polar_dis} shows that although the trend 
versus $U$ is 
the same in both supercell and bulk calculations, the 
magnitudes of $\Delta E$ do not agree at all. There must be something at 
the interface which is absent in bulk phase and 
significantly affects the energy difference between antiferro- and 
ferromagnetism. 

\begin{table}[t]
\caption{\label{tab:polar_dis} The comparison of energy difference 
between the interface calculations and bulk La$_{1-x}$Sr$_x$MnO$_3$ 
calculations. $\overline{x}=(x_1+x_2)/2$ is the average hole in the 
first and second La$_{1-x}$Sr$_x$MnO$_3$ layers closest to the interface. 
For different Hubbard $U$, $\overline{x}$ does not change significantly. 
$\Delta E_{I}$ is the energy difference between phases $A1$ and $F$ from the 
supercell calculations.  $\Delta E_{B}$ is the energy difference of 
SrTiO$_3$-strained \textit{bulk} 
\lsm~between $A$-type antiferromagnetism ($A$) and ferromagnetism ($F$).
$\Delta E_{BP}$ is the energy difference between $E(A)$ and $E(F)$ 
of SrTiO$_3$-strained \textit{bulk} \lsm, with the interfacial 
polar distortions 
manually included and $c$-axis optimized. For bulk calculations 
($\Delta E_{B}$ and $\Delta E_{BP}$), the nominal 
hole density is chosen as $\overline{x}$. }
\begin{center}
\begin{tabular}{c|c|c|c|c}
\hline
\hline 
$U$ (eV) & $\overline{x}$ & $\Delta E_{I}$ (meV) & $\Delta E_{B}$ (meV) & $\Delta E_{BP}$ (meV) \\
\hline
1        &   0.7          &   -34            &      -112         & -47                \\  
\hline
2        &   0.7          &    -8             &       -70         & -17                \\
\hline
3        &   0.65         &    14           &       -24         &  16                \\
\hline
4        &   0.6          &    42            &        19         &  57                \\    
\hline
\hline
\end{tabular}
\end{center}
\end{table}

\begin{figure}[t]
\includegraphics[angle=0,width=12cm]{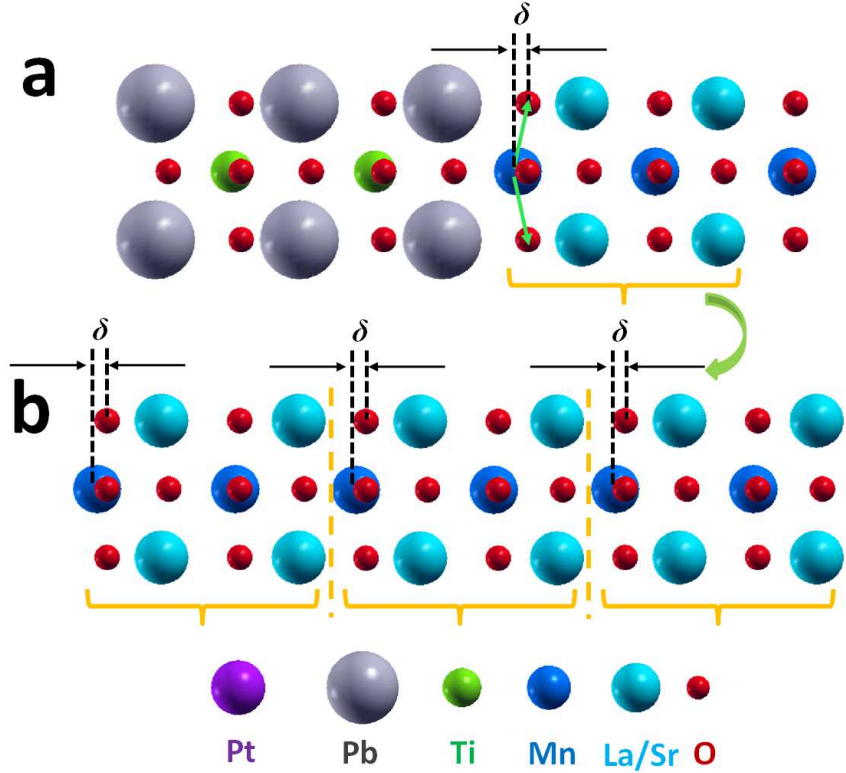}
\caption{\label{fig:polar_distortion} (Color online) \textbf{a)} Strong polar distortion is 
induced at the interface layer of MnO$_2$ due to the presence of PbTiO$_3$.
\textbf{b)} Schematics of copying the two interfacial \lsm~layers and 
forming artificial bulk La$_{1-\overline{x}}$Sr$_{\overline{x}}$MnO$_3$ with 
built-in polar distortions.}
\end{figure}

We find that, due to the presence of ferroelectric PbTiO$_3$, 
strong polar distortions are induced at the interface layer of MnO$_2$ in 
the accumulation state (Table~\ref{tab:distortions}), 
as is illustrated in Fig. \ref{fig:polar_distortion}a. 
The cation-oxygen vertical
displacement in MnO$_2$ at the interface is $\delta = 0.2$~\AA. 
Such a strong polar distortion (distinguished from the structural distortions 
of $Pnma$ symmetry) 
is absent in bulk \lsm~and is a direct consequence of the 
ferroelectric/manganite interface. In order to show explicitly 
how this interfacial polar distortion affects the 
energy difference between $F$ and $A$-type magnetic orderings, 
we perform the following thought experiment, which is schematically 
illustrated in Fig. \ref{fig:polar_distortion}b. 
We focus on the two layers of \lsm~closest to the interface in 
interface calculations, use their relaxed atomic positions and choose 
an $x$ that is equal to the average hole doping 
$\overline{x}=(x_1+x_2)/2$ from interface calculations. 
In this way, we create such artificial 
La$_{1-\overline{x}}$Sr$_{\overline{x}}$MnO$_3$ with the same 
built-in polar distortions and the same average hole doping as the 
two manganite 
layers at the interface. We impose periodic boundary conditions on this 
artificial La$_{1-\overline{x}}$Sr$_{\overline{x}}$MnO$_3$, fix all atom positions 
and in-plane lattice constants, and optimize the $c$-axis 
to minimize the out-of-plane stress. We tune $c$ separately 
for both ferromagnetism and $A$-type antiferromagnetism. A similar  
energy difference $\Delta E_{BP}$ (subscript $P$ means `polarized') 
is defined as 

\begin{equation}
\label{equ:ediffBP}\Delta E_{BP}=\frac{1}{2}\left(E_{BP}(A)-E_{BP}(F)\right)
\end{equation} 
where $E_{BP}(F)$ and $E_{BP}(A)$ are the total energies of 
the artificially constructed La$_{1-\overline{x}}$Sr$_{\overline{x}}$MnO$_3$ with 
ferromagnetic and $A$-type antiferromagnetic ordering, respectively. 
We can see from Table~\ref{tab:polar_dis} that $\Delta E_{BP}$ is 
much closer to $\Delta E_{I}$ than the raw bulk data $\Delta E_{B}$, 
demonstrating that 
in order to quantitatively connect the phase evolution of the 
ferroelectric/manganite interface from the bulk manganite phases, 
the polar distortion induced in interfacial manganites is an essential 
ingredient in modelling.

\section{Conclusion}
\label{conclusion}

We have presented a systemic study of the PbTiO$_3$/\lsm~ interface as a 
prototype for ferroeletric/manganite interfaces.  We are able to show
that the screening charges produced in the manganite in response to the
ferroelectric surface charge are sufficient to change the magnetic state
of the interfacial manganite from ferromagnetic to antiferromagnetic, in
agreement with experimental observations and interpretations. In the 
process, we have developed a method to accurately count the layer-by-layer 
hole distribution in the manganite thin film which allows us to perform
quantitative analysis of the system.  For example, it allows us to create
a simple Ising-like model of the interfacial magnetism that uses bulk
parameters to reproduce the computed behaviors.  

One of main theoretical findings is that the ground-state magnetic
state depends sensitively on the value of $U$ chosen in the LSDA+$U$
computation.  We show that different reasonable-seeming approaches to
determining $U$, and in particular some that are {\it ab initio} and
deliver a $U$ value appropriate to LSDA+$U$ self-consistently, yield
significantly different $U$ values. Not all the values do well when
compared to experiment.  By asking that the LSDA+$U$ calculation should
correctly reproduce the critical hole doping density separating the
ferromagnetic and antiferromagnetic phases, we are able to find a
narrow range of $U$ values that also produce a straightforward
interfacial magnetic ground-state structure (the $A1$ configuration)
whereby the manganite layer with the highest doping has the
strongest magnetic response. Clearly, our conclusions on the magnetic
ground-state are not {\it ab initio} as they involve significant
experimental input.  In our opinion, the unsatisfactory situation vis
a vis choosing the $U$ value is due to the limitations of the
single-particle DFT+$U$ method itself when applied to a complex and
strongly correlated electronic system such as manganites: the
theory is not accurate enough for the material, so some level of
empiricism is unfortunately necessary.

\begin{acknowledgments}
We are grateful to useful discussions with Carlos A. F. Vaz, Jason Hoffman, 
Yaron Segal, Fred J. Walker, Alexie M. Kolpak and Charles H. Ahn. 
This work was supported primarily by the National Science
Foundation under Contracts No. MRSEC DMR 0520495 and DMR 1119826 and 
in part by the facilities and staff of the Yale University Faculty of 
Arts and Sciences High Performance Computing Center and 
by the National Science Foundation under grant \#CNS 08-21132 that 
partially funded acquisition of the facilities.
Bulldog parallel clusters of the Yale High Performance Computing 
center and TeraGrid provided computational resources. 
\end{acknowledgments}

\appendix

\section{Tests of the virtual crystal approximation}
\label{virtualxtal}

We treat the $A$-site La$_{1-x}$Sr$_{x}$ alloying in \lsm~with the virtual 
crystal approximation \cite{Nordheim-AnnPhys-1931, Vanderbilt-PRB-2000}.
This approximation involves replacing the two elements by a fictitious
one whose electron number is $(1-x)N_{\textrm{La}}+
xN_{\textrm{Sr}}$, where $N_{\textrm{La}}$ and $N_{\textrm{Sr}}$ are 
the number of electrons of the La and Sr pseudo atoms, respectively.
We stress that i) since the magnetic properties originate 
from Mn $d$-electrons and $A$-site atoms serve to donate 
electrons, we expect that the virtual crystal approximation is reasonable 
to describe the magnetic phase transition of manganites in the random 
distribution; ii) since the chemical properties mainly depend on the valence 
electrons, the approximation we make here is expected to be also 
good for Ca and Ba, and iii) our 
choice of pseudo potential and valence electrons shown in Table 
\ref{tab:psp} ensures a very smooth interpolation between La and Sr as 
their pseudo valence electrons and nuclear pseudo charges only 
differ by one elementary charge. We perform simple tests of 
1:1 Sr-La alloying (i.e. 50\% alloying) in a $c(2\times2)\times2$ unit cell and the
results are compared to the $x=0.5$ virtual 
crystal approximation as shown in Table~\ref{tab:vca}.
We can see that virtual crystal approximation quantitatively 
reproduces the lattice constants and the energy differences 
between various magnetic orderings when compared to the calculation
with ``real'' La and Sr atoms. The accuracy of magnetic energy differences
is consistent with earlier work
~\cite{Fang-PRL-2000}. 

\begin{table}[h]
\caption{\label{tab:vca} Comparison between the virtual crystal approximation 
and supercell calculations. The nominal doping $x$ is 0.5 in the virtual 
crystal approximation. A $c(2\times2)\times 2$ supercell is employed with 
La and Sr atoms forming a checker-board pattern (every nearest neighbor of 
Sr is La and vice versa). The lattice constants reported are 
those for $A$-type antiferromagnetic ordering. $\Delta E$ is the energy 
difference between 
ferromagnetic ordering and $A$-type antiferromagnetic ordering  
per Mn atom, defined by Eq.~(\ref{equ:deltaE}). A range of 
Hubbard $U$ (0 $\leq U \leq 2$ eV) 
are tested.}
\begin{center}
\begin{tabular}{c|c|c|c|c|c|c|c|c|c|c|c|c|c|c}
\hline
\hline
 & \multicolumn{7}{c|}{Virtual crystal approximation} & \multicolumn{7}{c}{$c(2\times2)\times 2$ supercell} \\
\hline    
$U$ (eV)  & \multicolumn{2}{c|}{$a$ (\AA)}  & \multicolumn{2}{c|}{$b$ (\AA)} & \multicolumn{2}{c|}{$c$ (\AA)} & $\Delta E$ (meV)& \multicolumn{2}{c|}{$a$ (\AA)}  & \multicolumn{2}{c|}{$b$ (\AA)} & \multicolumn{2}{c|}{$c$ (\AA)} & $\Delta E$ (meV) \\  
\hline
 & $F$ & $A$ & $F$ & $A$ & $F$ & $A$ &  & $F$ & $A$ & $F$ & $A$ & $F$ & $A$ & \\ 
\hline
0   & 5.345 & 5.363  & 7.561 & 7.385  & 5.386 & 5.408  & -34  & 5.327 & 5.366  &  7.482 & 7.372 & 5.381 & 5.411 &  -37 \\
\hline
1   & 5.385 & 5.376  & 7.612 & 7.411  & 5.430 & 5.433  & -10  & 5.355 & 5.380  & 7.513 & 7.398 & 5.400 & 5.439 &  -11 \\  
\hline
2   & 5.401 & 5.380  & 7.623 & 7.422  & 5.434 & 5.452  &  7 & 5.365  & 5.387  & 7.520 & 7.411 & 5.410 & 5.452 &   6  \\
\hline
\hline
\end{tabular}
\end{center}
\end{table}

\section{The effects of SrTiO$_3$ substrate}
\label{sto}

\begin{figure}[t]
\includegraphics[angle=-90,width=12cm]{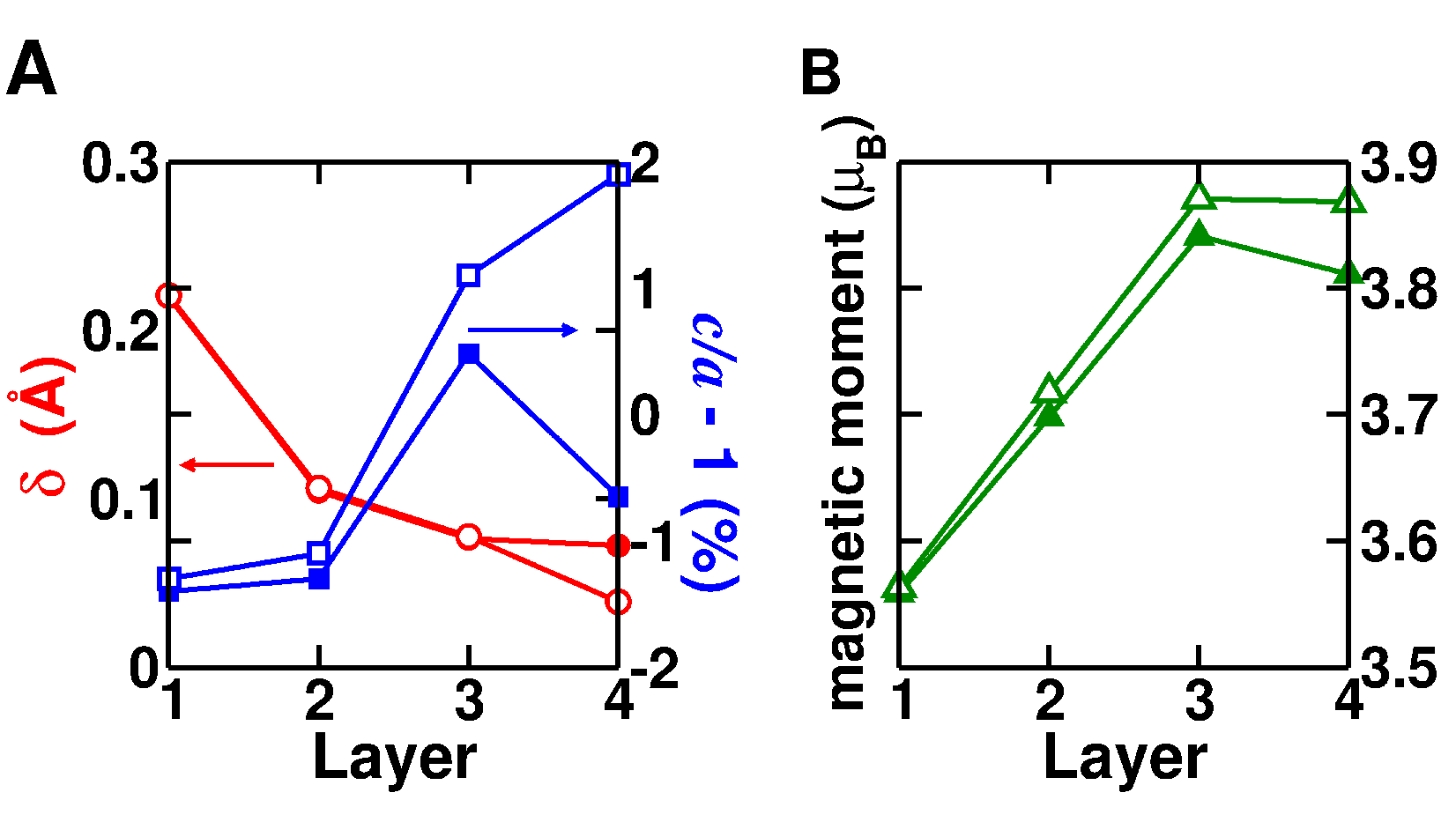}
\caption{\label{fig:sto_effect} (Color online) The effects of SrTiO$_3$ substrate on 
structural and magnetic properties at the PbTiO$_3$/\lsm~interface. Layer1 
is the interfacial layer of MnO$_2$.
The solid symbols correspond to the data with SrTiO$_3$ substrate. 
The open symbols correspond to the data without SrTiO$_3$ substrate. 
\textbf{A)} rumplings of each MnO$_2$ layer and $c/a$ ratio of each oxygen 
octahedron that encloses Mn atoms. \textbf{B)} $d$-orbital magnetic 
moment of each Mn atom, calculated by using L\"{o}wdin orbitals.}
\end{figure}

Inclusion of SrTiO$_3$ substrate in the simulation of PbTiO$_3$/\lsm~
heterostructures is computationally intensive. In this section, we test 
the effects of SrTiO$_3$ substrate on structrual and magnetic properties 
on the PbTiO$_3$/\lsm~interface. 
We compare two calculations: one with three unit cells 
of SrTiO$_3$ and the other without SrTiO$_3$. \lsm~is four unit cells thick 
with nominal doping $x=0.2$. PbTiO$_3$ thin film is polarized so that both 
calculations are in the accumulation state. In terms of structural 
properties, we focus on two important quantities. One is the $c/a$ ratio 
of each oxygen octahedron that encloses Mn atoms. The other is the rumpling 
$\delta$ of each MnO$_2$ layer. The results are shown in 
Fig.~\ref{fig:sto_effect}A. 
Layer 1 is the interface. Layer 4 is the artificial surface when SrTiO$_3$ 
is absent. The solid symbols are with SrTiO$_3$ and the open symbols 
are without SrTiO$_3$. We can see that the structural properties with or 
without SrTiO$_3$ substrate quickly converge as the interface is approached. 
At Layer 1, the difference is negligible. In terms of magnetic properties, 
we calculate the $d$-orbital magnetic moment of each Mn atom, using the 
L\"{o}wdin orbitals approach~\cite{Lowdin-JChem-1950}. 
Fig.~\ref{fig:sto_effect}B shows the comparison between with SrTiO$_3$ 
substrate (solid symbols) and without SrTiO$_3$ substrate (open symbols). 
Similar to structural properties, the effects of SrTiO$_3$ substrate on 
magnetic properties are generally very small and diminish at the interface.
Therefore we do not include SrTiO$_3$ substrate in our simulation, 
not only to reduce the computation burden but also to introduce an 
artificial surface so that we can apply a counting method 
(see Appendix~\ref{count}) to accurately calculate hole distribution in \lsm.

\section{The method for counting holes}
\label{count}

\begin{figure}[t]
\includegraphics[angle=0,width=12cm]{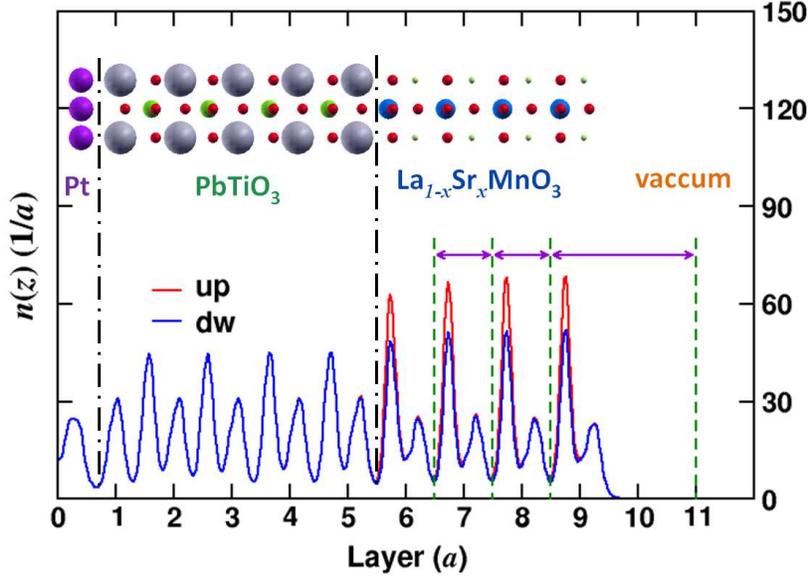}
\caption{\label{fig:charge_profile}(Color online) Illustration of how to count the charge 
in each layer of \lsm. The spin polarized part is \lsm. The green dashed 
lines highlight the computed boundaries of each layer.}
\end{figure}

A widely used approach for calculating the number of holes and 
the magnetization of Mn atoms is to 
use L\"{o}wdin orbitals~\cite{Lowdin-JChem-1950}. 
However, a more direct method is 
to use the electron density itself. The difficulty lies in 
that the boundary between each manganite layer is not well-defined 
in the thin film of \lsm. We develop a method to 
self-consistently set the boundary between each manganite layer, provided 
that the manganite is half-metallic. 

For a half-metallic manganite, there are no states at the Fermi level 
in the minority spin channel, so that there must be a definite integer number
of electrons $N_c$ filled in the minority spin channel. $N_c$ depends 
on the details of pseudopotentials. For our 
pseudo atoms (see Table \ref{tab:psp}), for doping level $x$, 
we have La$^{3+}$: $5s^25p^65d^06s^0$, 
Sr$^{2+}$: $4s^24p^65s^0$, O$^{2-}$: $2s^22p^6$ and due to charge 
conservation, Mn ion is nominally $+(3+x)$ with an electron 
configuration $3s^23p^63d^{4-x}4s^{0}$. For one unit cell of \lsm,
since spin polarization only comes from the electrons on Mn $d$-orbitals, 
we can sum all the other electrons that are formally spin unpolarized:
$8\times(1-x)+8\times x+8\times3+8=40$. The four terms are from La$^{3+}$,
Sr$^{2+}$, three O$^{2-}$ and the Mn ion without $d$-electrons. 
Hence $N_c=40/2=20$, which is independent of hole doping. Now
we start from the vacuum (see Fig. \ref{fig:charge_profile}) 
where there is no charge. We 
integrate the minority spin channel moving into the film until the integral 
is equal to 20. Then this position determines the boundary of the 
first layer. 
Next we restart the integral from this boundary until it reaches 20 
again. This determines the boundary of the second layer. Repeating 
the procedure yields the boundaries of each manganite layer. Once 
the boundaries are determined, we integrate the charge density of 
both majority and minority spins in each layer and thus layer-resolved 
holes and magntization follow straightforwardly.   

We comment that in DFT calculations, 
as long as the Hubbard $U$ is larger than a critical value $U_c$, the 
manganites become half-metallic in the ferromagnetic phase. For 
SrTiO$_3$-strained $Pnma$ \lsm, we find $U_c\simeq 1$ eV. Therefore, 
for the useful and reasonable range of $U$, our method is valid.

\section{The phase diagram of manganites from LSDA+$U$}
\label{fullphase}

\begin{figure}[t]
\includegraphics[angle=-90,width=12cm]{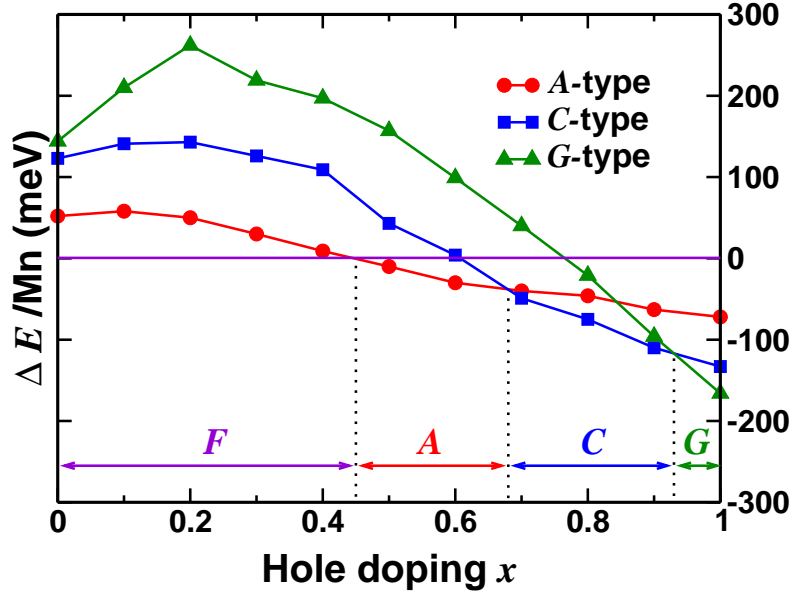}
\caption{\label{fig:full_phase} (Color online) The complete magnetic phase diagram of 
\lsm~(4 formula, 20 atoms). The calculation is based on LSDA+$U$ with 
$U$ = 1 eV. $\Delta E$ is the energy difference between ferromagnetism and 
various types of anti-ferromagnetism. 
The labels $F$, $A$, $C$ and $G$ refer to ferromagnetism, 
$A$-type, $C$-type and $G$-type antiferromagnetism, respectively. Each 
label highlights the ground state magnetic structure of the given 
hole doping region.}
\end{figure}

In this section, we provide the complete magnetic phase diagram of 
La$_{1-x}$Sr$_x$MnO$_3$ in Fig.~\ref{fig:full_phase} 
based on LSDA+$U$ with $U$ = 1 eV. In the high doping region ($x > 0.4$), 
LSDA+$U$ does reproduce the experimentally observed 
sequence of different magnetic 
ground states~\cite{Fang-PRL-2000}: 
$\textrm{FM}\to A-\textrm{AFM}\to C-\textrm{AFM}\to G-\textrm{AFM}$ 
as the hole doping $x$ 
increases. Our result is consistent with previous calculations
~\cite{Burton-PRB-2009}.

\section{Test of band alignment and possible artificial charge spillage}
\label{bandalign}

\begin{figure}[t]
\includegraphics[angle=-90,width=12cm]{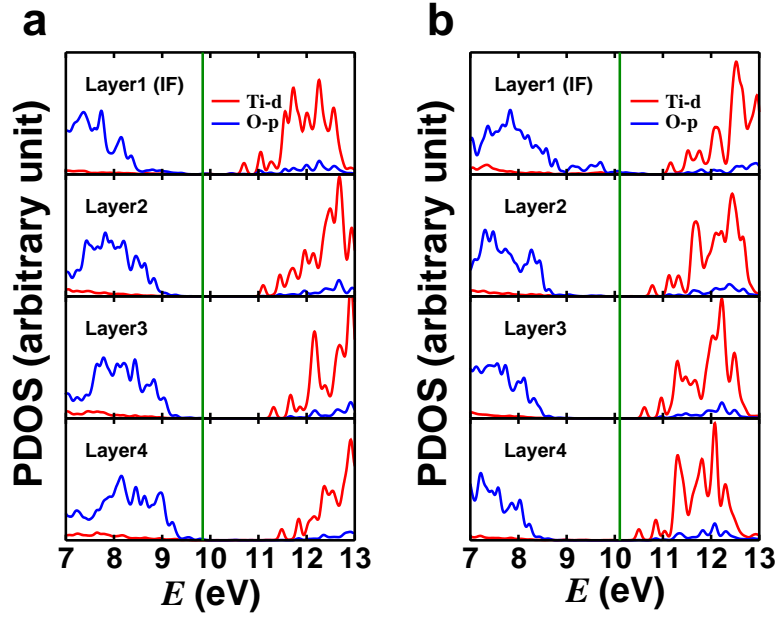}
\caption{\label{fig:band_align} (Color online) Atomic projected density of states (PDOS) of 
PbTiO$_3$ in the PbTiO$_3$/La$_{1-x}$Sr$_x$MO$_3$/Pt heterostructures. 
\textbf{a)} The accumulation state. 
\textbf{b)} The depletion state. The red curves are Ti-$d$ projected states 
and the blue curves are O-$p$ projected states. Layer1 refers to the interface 
between PbTiO$_3$ and \lsm~(the interface 
we are interested in) and Layer4 is the interface between PbTiO$_3$ 
and Pt. The green solid line is the Fermi level. }
\end{figure}

Due to the well known underestimation of band gaps in DFT
calculations, band alignment errors and possible artificial charge
spillage into the conduction bands of on material at an interface may
occur.  These errors can lead to unrealistic ground states when
simulating the interface between ferroelectrics and
metals~\cite{Stengel-PRB-2011}.  We check our calculations of
PbTiO$_3$/La$_{1-x}$Sr$_x$MO$_3$/Pt interface and find that the Fermi
level is in the band gap of PbTiO$_3$. A typical projected density of
states (PDOS) of both the accumulation and depletion states 
is illustrated in Fig.~\ref{fig:band_align}, where
Layer1 refers to the interface between PbTiO$_3$ and \lsm~ (the
interface we are interested in) and Layer4 is the interface between
PbTiO$_3$ and Pt. We can see that the interior of PbTiO$_3$ remains
insulating. We need to point out that both terminations of PbTiO$_3$
are PbO layers in our calculations, instead of the pathological
TiO$_2$ termination which leads to a metallic ferroelectric ground
state in other similar systems~\cite{Stengel-PRB-2011}.

\end{document}